\documentclass[journal,onecolumn,12pt]{IEEEtran}
\usepackage{epsfig,rotating,setspace,latexsym,amsmath,epsf,amssymb,bm,theorem}
\usepackage{cite,psfrag,bbm}
\usepackage{amsfonts,authblk,subfigure}
\usepackage{graphicx}
\newtheorem{Lemma}{Lemma}

\def \n2{{N_0 \over 2}}

\def \h5{\hspace{0.5in}}

\begin{document}

\title{Waiting before Serving: A Companion to Packet Management in Status Update Systems}

\author{Peng Zou \qquad Omur Ozel \qquad Suresh Subramaniam\thanks{The authors are with the Department of Electrical and Computer Engineering, George Washington University, Washington, DC 20052. Emails: \{pzou94, ozel, suresh\}@gwu.edu. Part of this work appears in the Proceeding of IEEE INFOCOM Age of Information Workshop, Paris, France. April 2019.}}

\maketitle
\vspace{1.0cm}
\begin{abstract}
In this paper, we explore the potential of server waiting before packet transmission in improving the Age of Information (AoI) in status update systems. We consider a non-preemptive queue with Poisson arrivals and independent general service distribution and we incorporate waiting before serving in two packet management schemes: M/GI/1/1 and M/GI/1/$2^*$. In M/GI/1/1 scheme, the server waits for a \textit{deterministic} time immediately after a packet enters the server. In M/GI/1/$2^*$ scheme, depending on idle or busy system state, the server waits for a \textit{deterministic} time before starting service of the packet. In both cases, if a potential newer arrival is captured existing packet is discarded. Different from most existing works, we analyze AoI evolution by indexing the incoming packets, which is enabled by an alternative method of partitioning the area under the evolution of instantaneous AoI to calculate its time average. We obtain expressions for average and average peak AoI for both queueing disciplines with waiting. Our numerical results demonstrate that waiting before service can bring significant improvement in average age, particularly, for heavy-tailed service distributions. This improvement comes at the expense of an increase in average peak AoI. We highlight the trade-off between average and average peak AoI generated by waiting before serving. 
\end{abstract}

\pagestyle{plain}
\pagenumbering{arabic}

\newpage

\section{Introduction}

Age of Information (AoI) is a metric measuring the staleness of available information at the receiver of a system that monitors a physical phenomenon of interest and updates the status. Since its early treatments in \cite{kaul2012status, kaul2012real} for queuing models motivated from vehicular status update systems, the AoI metric has been found useful in and related to numerous applications that require timely availability of information at the receiving end of a communication system. In particular, \cite{costa2016age, kam2018age} investigates the role of packet management with the possibility of packet deadlines to improve the AoI at the monitoring node. \cite{inoue2018general} provides a general treatment of stationary probability analysis of AoI in various preemptive and non-preemptive queuing disciplines; see also \cite{najm2016age, najm2017status} for more specialized studies. \cite{2018information} provides an information-theoretic treatment of the tradeoff between AoI and throughput in an energy harvesting timing channel. We also refer to \cite{bacinoglu2015age, yates2015lazy, wu2017optimal_ieee, arafa2017age, arafa2017age2} for AoI in energy harvesting communication systems. \cite{wang2018skip} considers AoI under link capacity constraints and \cite{kosta2017age2} considers non-linear age dimension into the problem. Evolutions of AoI through multiple hops in networks have been characterized in \cite{bedewy2017minimizing, bedewy2017age, bedewy2017age2, yates2018age, yates2018status, maatouk2018age}. References \cite{hsu2017age, talak2017minimizing, jiang2018decentralized} consider AoI optimization over broadcast and multi-access scenarios.

In this paper, we consider a point-to-point status update system where the transmitting node sends status updates to the receiving node through a queue. In this abstraction, the status update age is the time elapsed since the last received sample was generated. The content of the message is assumed irrelevant in the formulation. We investigate average AoI and average peak AoI where randomly generated samples arrive according to a Poisson process and the time it takes for a packet to be transmitted has a general probability distribution. At this point, we bring the seminal paper \cite{sun2017update} into attention. In this reference, general insights and analysis are provided as to when ``waiting before updating" is useful to improve AoI performance in a point-to-point status update system. In the setting of \cite{sun2017update}, the samples are generated one at a time at the source in the presence of {\em perfect knowledge} of the server state. It has been shown analytically and numerically that heavy-tailed service distributions are especially amenable to provide cases of boosted AoI when a deliberate waiting period is introduced in the status update generation process. Our current paper explores the benefits of waiting further. However, our system is different from~\cite{sun2017update} in that status update packets are generated at random times one after the other {\em independently} in our model. We assume no feedback of the server state, and packet generation is oblivious to the transmitter state. Additionally, we allow the transmitter to manage packet transmissions by discarding earlier updates when a later update arrives at the transmitter, and introducing a delay before transmitting an available status update. Through decoupling the update generation process from the communication process, we aim to capture a natural characteristic of various types of applications in which sensors generate updates oblivious to server state and the service of the status update packets is separately handled. 

Among earlier works, references \cite{bedewy2017minimizing, bedewy2017age, bedewy2017age2, sun2018age} consider a special class of service distributions (termed New Better than Used (NBTU)) with Poisson arrivals in the context of packet management for single and multihop cases. These references prove that under such service distributions M/GI/1/$2^*$ scheme (or last come first serve with preemption only in waiting as referred in these references) is near optimal with a constant gap to optimality. Our results are derived under a general service distribution and we allow server to perform additional waiting (in contrast to a work conserving scheduling policy). Our work significantly extends the understanding in this direction through determining gains obtained by ``server waiting'' in various cases and comparing packet management schemes with waiting. 

In this paper, we consider M/GI/1/1 and M/GI/1/$2^*$ queuing disciplines compatible with Kendall notation, reminiscent of the one used in \cite{costa2016age} (M/GI/1/$2^*$ is called in \cite{inoue2018general} M/GI/1 with last come first serve and discarding). We consider modified versions of these queuing disciplines through introducing additional waiting before serving. In both schemes, the updates arrive at the transmitter according to a Poisson process, and the time it takes for a packet to be transmitted is a random variable that has a general distribution, is independent over time, and is also independent of other events in the system. In M/GI/1/1 scheme, there is no data buffer to store incoming packets and an arrival is taken to service only if the server is not serving another packet. In M/GI/1/$2^*$ scheme, a single buffer is available in the queue, so the system can store one packet while the server is busy. \textit{In both schemes, the server is not equipped with the option to preempt service for a new arrival.} Instead, the server \textit{waits} an additional time before continuing to serve the latest arriving packet. In M/GI/1/1 scheme, this waiting happens once a packet enters the server after an idle period. In M/GI/1/$2^*$ scheme, this waiting happens once a packet enters the server after an idle period or a busy period. Finally, we assume that any packet in the buffer is discarded if a new update arrives while the server is busy or waiting before service. \textit{The potential benefit expected from waiting is to capture newer packets at the expense of longer wait times for packets in service.} We allow deterministic amounts of waiting after idle and busy periods of the server, and perform stationary distribution analysis to obtain expressions for average AoI.

We determine closed form expressions for average AoI and average peak AoI for M/GI/1/1 and M/GI/1/$2^*$ schemes with waiting. We obtain numerical results by evaluating the expressions we find. Our numerical results demonstrate that waiting is especially helpful for heavy-tailed service distributions such as inverse Gaussian distribution while the improvement is limited for light-tailed ones such as exponential and Erlang distributions. With regard to the comparison between M/GI/1/1 and M/GI/1/$2^*$, our results show that  the latter outperforms the former once waiting is optimized. Still, the improvement brought by the presence of a data buffer appears to be small. This motivates further studies on understanding if use of a buffer space can be exchanged with waiting before serving in a single server status update system. Additionally, we observe that the improvement brought by ``server waiting" comes at the expense of an increase in average peak AoI and we obtain the optimal tradeoff curves achieved by deterministic waiting schemes. Our numerical findings highlight the trade-off between average and average peak AoI generated by waiting before serving.

\section{The Model}
\label{sec:Model}

We consider a point-to-point link with a single transmitter (or server) and a single receiver. The status updates arrive to the transmitter in a packet form according to a Poisson process of rate $\lambda$. The transmitter node transmits the status update packets one at a time. The time for a packet to be served is independent of other system variables and independent for each packet with a general service time density function $f_S(s)$, $s \geq 0$. We use $MGF^{(S)}_{\gamma}$ to denote the moment generating function of the service distribution evaluated at $-\gamma$:
\begin{align}
MGF^{(S)}_{\gamma} = \mathbb{E}[e^{-\gamma S}]
\end{align}
where we are interested in $\gamma \geq 0$. To use in the ensuing analysis, we also define the following:
\begin{align}\label{eq:def}
MGF_{\gamma}^{(S,1)}\triangleq \mathbb{E}[Se^{-\gamma S}], \ MGF_{\gamma}^{(S,2)} \triangleq \mathbb{E}[S^2e^{-\gamma S}]
\end{align}
where $MGF_{\gamma}^{(S,1)}$ and $MGF_{\gamma}^{(S,2)}$ are the first and second derivative of the moment generating function of $S$ at $-\gamma$. We use $t_i$ to denote the time stamp of the event that packet $i$ enters the queue, and $t_i'$ to denote the time stamp of the event that the service of packet $i$ (if selected for service) is completed and it is delivered to the receiver. 

We consider two packet management schemes M/GI/1/1 and M/GI/1/$2^*$. In M/GI/1/1 scheme, there is no data buffer and a packet is accepted to the server if it is idle. In M/GI/1/$2^*$ scheme, we assume that a single packet may be kept in queue. The transmitter sends the latest arriving update and discards the previous updates. With Poisson arrivals and general service time, a single server, and one space in the buffer, this model of packet management is in the form of an M/GI/1/$2^*$ queue, referring to the usual M/GI/1/2 with the additional modification due to packet discarding. 

\subsection{Deterministic Waiting Policy}

In both queuing schemes, the server waits before starting service in the same spirit as \cite{sun2017update}. In M/GI/1/1 scheme, packets arriving when the server is busy are discarded, and packets arriving in the idle state wait at the server for $\epsilon_I$ duration before service starts. If a new arrival occurs in this duration, the existing packet in the server is discarded and is replaced with the new one. In M/GI/1/$2^*$ scheme, a packet arriving to an idle system is treated exactly the same as in M/GI/1/1. A packet arriving to a busy system is stored in the buffer (replacing the packet in the queue, if there is already one). When service of the current packet ends, we introduce an additional waiting, which we term waiting after a busy period of a deterministic amount $\epsilon_{B}$. If during this $\epsilon_{B}$ period a new packet arrives, the new arrival is served and the existing one is discarded at the end of the ongoing waiting period. We assume that the waiting times are decided beforehand and are applied invariantly throughout the process. In both queuing disciplines considered in this paper, the instantaneous Age of Information (AoI) is measured by the difference of the current time and the time stamp of the latest delivered packet at the receiver: 
\begin{align}
\Delta(t)=t-u(t)
\end{align}
where $u(t)$ is the time stamp of the latest received packet at time $t$.

\subsection{Equivalent Queuing Model for M/GI/1/1}
\label{sec:eqmodel1}

We now present another queuing model for M/GI/1/1 scheme that yields an identical AoI pattern to our system's, and one we will use to analyze the average and average peak AoI in our system. In this model, the data buffer capacity is unlimited. Each arriving packet is stored in the queue and no packet is discarded. We allow multiple packets to be served at the same time. An arriving packet may find the system in three different states: i) Idle (I), ii) Busy (B), and iii) Waiting (W). If a packet finds the system in state (I), then that packet's service starts after $\epsilon_{I}$ units of time together with all other packets that arrive during this waiting period finding the system in state (W). If an arriving packet finds the system in state (B), its service starts after the end of the current service period, the idle period waited for the arrival of the next packet and the additional waiting period $\epsilon_{I}$. The packets arriving to the system in state (B) are served together with the next arriving packet as well as all other packets that arrive during the following (W) period.    

We note that this equivalent queue model is not physically the same as the original model in that the queue can hold at most one packet and the server cannot serve multiple packets simultaneously in the original model. Nevertheless, this queuing model yields an AoI evolution over time that is identical to the original non-preemptive M/GI/1/1 model. With this new queuing, we essentially allow the discarding of the incoming packets to happen at the end of the ensuing service time for the next arrival and this has no influence on the AoI evolution. In this equivalent model, no packet is discarded and this enables us to index the arriving packets.  

\begin{figure}[!t]
\centering{
\hspace{-0.2cm} 
\includegraphics[totalheight=0.32\textheight]{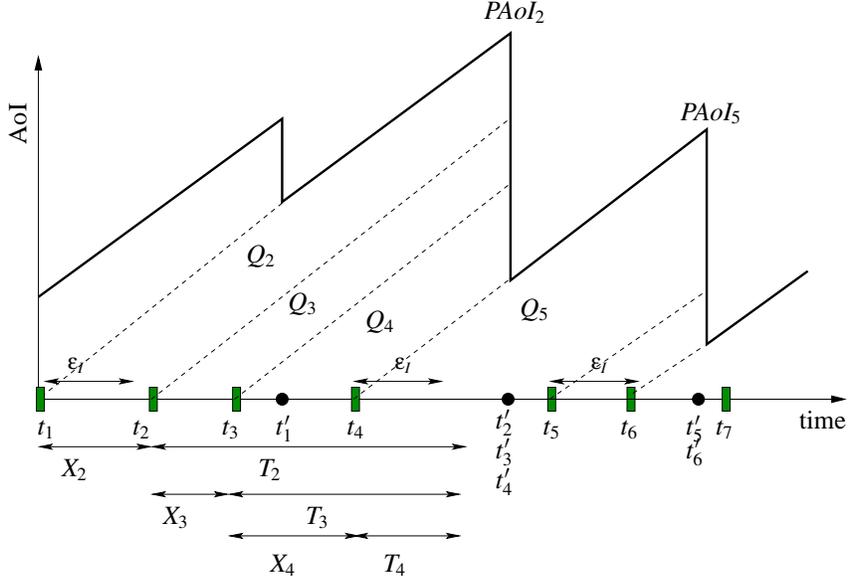}}\vspace{-0.1in}
\caption{\sl Example AoI evolution for the equivalent queuing model of M/GI/1/1 with waiting.}
\label{fig:1} 
\vspace{-0.2in}
\end{figure}

We provide an instantiation of the AoI evolution under this equivalent queuing model in Fig. \ref{fig:1}. We assume packet 1 finds the server idle and the server waits $\epsilon_I$ time units to capture an incoming packet. During the service of packet $1$ in between $t_1$ and $t_1'$, packets 2 and 3 arrive and they are both saved in the buffer. Note that in the original model, these packets are discarded. The queue is idle at $t_1'$ and the server waits until packet 4 arrives when a waiting period of $\epsilon_{I}$ duration starts. The packets 2, 3 and 4 are served simultaneously at the end of the waiting period. Note that the end of service times $t_2'$, $t_3'$, $t_4'$ coincide as shown in Fig. \ref{fig:1}. In the original model, only packet 4 is served. At time $t_4'$, the system enters idle state, and packet 5 finds the system idle. At $t_5$, the system enters (W) state for $\epsilon_{I}$ time and during this period packet 6 arrives. Both packets 5 and 6 are served together and their services end at coincident times $t_5'$, $t_6'$. In the original model, packet 5 is discarded and only packet 6 is taken into service.

\subsection{Equivalent Queuing Model for M/GI/1/$2^*$}
\label{sec:eqmodel2}

We now present an equivalent queuing model for M/GI/1/$2^*$. In this model, the data buffer capacity is unlimited, no packet is discarded and multiple packets are served at the same time. An arriving packet may find the system in four different states: i) Idle (I), ii) Busy (B), iii) Waiting after an Idle Period (WaI) and iv) Waiting after a Busy Period (WaB). If a packet finds the system in state (I), then that packet's service starts after $\epsilon_{I}$ units of time together with all other packets that arrive during this waiting period finding the system in state (WaI). If an arriving packet finds the system in state (B), its service starts after the end of the current service period and the additional waiting period $\epsilon_{B}$. The packets arriving to the system in state (B) are served together with all other packets that arrive during the same busy period and the following (WaB) period.

\begin{figure}[!t]
\centering{
\hspace{-0.2cm} 
\includegraphics[totalheight=0.32\textheight]{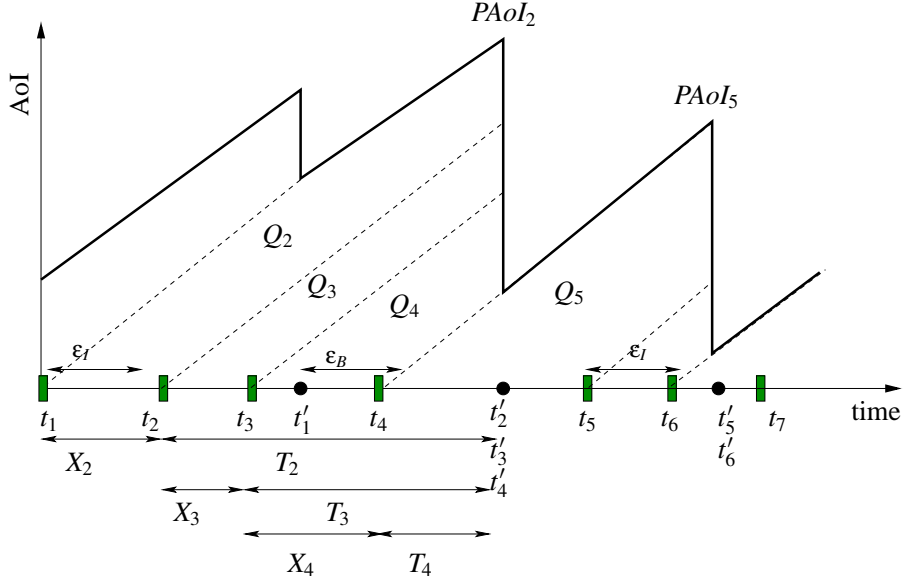}}\vspace{-0.1in}
\caption{\sl Example AoI evolution for the equivalent queuing model of M/GI/1/2* with waiting.}
\label{fig:3} 
\vspace{-0.2in}
\end{figure}

We present the AoI evolution under this equivalent queuing model in Fig. \ref{fig:3}. In this figure, the arrivals are identical to the one used in Fig. \ref{fig:1}. Packet 1 finds the server idle and an $\epsilon_I$ waiting period is added to capture a new arrival. The queue is not empty at $t_1'$ and so the server waits $\epsilon_{B}$ time units as the system state is busy at $[t_1']^{-}$. In this waiting period, packet 4 is captured and the packets 2, 3 and 4 are served simultaneously at the end of the waiting period. Note that the end of service times $t_2'$, $t_3'$, $t_4'$ coincide as shown in Fig. \ref{fig:1}. In the original model, only packet 4 is served and packets 2, 3 are discarded. At time $t_4'$, the system enters idle state, and packet 5 finds the system idle. At $t_5$, the server starts to wait $\epsilon_{I}$ duration and during this period packet 6 arrives. Both packets 5 and 6 are served together and their services end at coincident times $t_5'$, $t_6'$. 

For both M/GI/1/1 and M/GI/1/$2^*$ queueing disciplines, we define the areas $Q_i$ under the triangular regions of the AoI curve in the same order as the arriving packet indices as shown in Figs. \ref{fig:1} and \ref{fig:3}. These definitions are identical to those in \cite{kaul2012real} where first come first serve (FCFS) queuing is assumed. Note that the equivalent model in our current work is in the category of FCFS in that the packets are never discarded and they are served in the same order as they arrive. We now define $X_i$ as the length of time interval between the arrivals of packets $i-1$ and $i$ and $T_i$ as the system time for packet $i$ in the equivalent queuing model. These definitions are identical to those in \cite{kaul2012real}. We, therefore, have the average AoI as:
\begin{equation}\label{aaoi}
\mathbb{E}[\Delta]=\lambda\left(\mathbb{E}[XT]+\frac{\mathbb{E}[X^{2}]}{2}\right) = \lambda \mathbb{E}[XT] + \frac{1}{\lambda}
\end{equation}
More generally, the $k$th moment of AoI is as follows \cite{inoue2018general}
\begin{align}
\mathbb{E}[\Delta^k]=\lambda\left( \frac{\mathbb{E}[(X+T)^{k+1}] - \mathbb{E}[T^{k+1}]}{k+1} \right)
\end{align} 
We focus on the first moment in this paper. Extension to higher order moments can be made possible by following similar but more elaborate calculations. 

We also highlight the peak AoI evolution in Figs. \ref{fig:1} and \ref{fig:3}. In particular, $PAoI_{i^*}$ refers to the maximum $X_j+T_j$ among all packets $j$ served at a service period and $i^*$ refers to the packet index corresponding to the maximum. In these figures, packets $2$, $3$ and $4$ are served together and the maximum $X_j + T_j$ is $X_2 + T_2$. Similarly, packets $5$ and $6$ are served together and the maximum $X_j + T_j$ is $X_5 + T_5$. In general, it becomes apparent that the index that yields maximum $X_j + T_j$ is the minimum index among all that are served together. It is worthwhile to note that peak AoI is defined at the end of each service. Therefore, average peak AoI refers to an average over all service events rather than an average over time. On the contrary, AoI is defined at each time instant and average AoI represents average over the whole time duration. In particular, average AoI represents average of the area under AoI evolution while average peak AoI is average of just the edges appearing at peak points. This subtle fact determines the statistical behavior of average peak AoI and its comparison with average AoI. Since the system is ergodic, we will work with the generic variables for inter-arrival time $X$ and system time $T$. Similarly, we use $PAoI$ to denote the maximum $X_j + T_j$ among those that are served together. 

\section{Average AoI and Average Peak AoI for M/GI/1/$1$ with Waiting}
\label{sec:eval}

In this section, we consider average AoI and average peak AoI for M/GI/1/$1$ packet management with waiting.

\subsection{Average AoI}

In order to calculate average AoI, $\mathbb{E}[\Delta]$, in (\ref{aaoi}), it suffices to find the correlation between $X$ and $T$. Recall that $X$ is exponentially distributed with rate $\lambda$. As outlined in Section \ref{sec:eqmodel1}, the system can be in three different states. In view of the renewal structure, we have the following stationary probabilities for each state:
\begin{align}\label{one1}
p_{I} = \frac{1}{\lambda T_{cycle}}, \ p_{B}= \frac{\mathbb{E}[S]}{T_{cycle}}, \ p_{W}  = \frac{\epsilon_{I}}{T_{cycle}}
\end{align}
where $T_{cycle}$ is the expected length of one renewal cycle:
\begin{align}\hspace{-0.1in}
T_{cycle}=\frac{1}{\lambda}+\epsilon_{I}+\mathbb{E}[S]
\end{align}
These expressions are obtained by a standard application of Renewal Reward Theorem (see, e.g., \cite{gallager2012discrete,ross2014introduction}). In one renewal cycle in our queuing model, the system first starts in (I) state and shifts to (W) when an arrival occurs. The expected length of staying in (I) is $\frac{1}{\lambda}$. After staying in (W) for a deterministic $\epsilon_I$ time units, the system state switches to (B) and stays there for a service time. If an arrival occurs during service time, it is discarded. At the end of (B) period, the system then goes back to (I) and it completes one renewal cycle. By Renewal Reward Theorem, the stationary probabilities $p_s$, $s \in \mathcal{S}_{M/GI/1/1}=\{\mbox{I},\mbox{W},\mbox{B}\}$, in (\ref{one1}) are equal to the expected time spent in system state $s$ in one renewal cycle divided by the expected cycle length.  

\begin{figure}[!t]
\centering{
\hspace{-0.3cm} 
\includegraphics[totalheight=0.19\textheight]{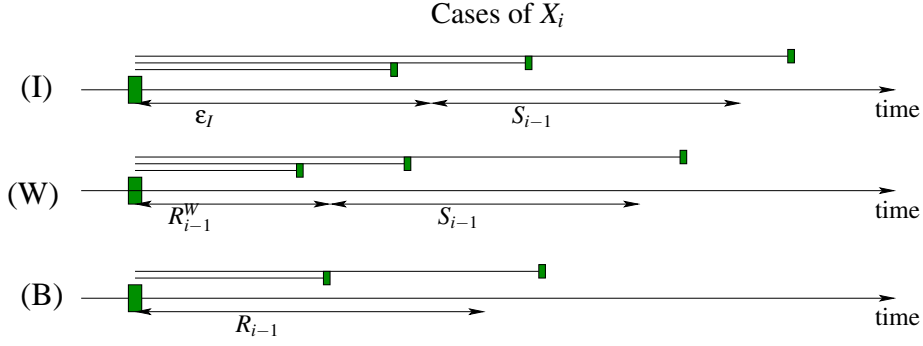}}\vspace{-0.1in}
\caption{\sl Three states of the system packet $i-1$ can observe in M/GI/1/1 scheme.}
\label{fig:4} 
\vspace{-0.2in}
\end{figure}

We next evaluate $\mathbb{E}[X_i T_i  \ | \ (s)]$ for $s \in \mathcal{S}_{M/GI/1/1}$ and conditioning is on the system state observed by packet $i-1$, denoted as $P_{i-1}$. Due to PASTA property, $\mbox{Pr}[P_{i-1}=(s)]=p_{s}$ where $p_s$, $s \in \mathcal{S}_{M/GI/1/1}$ are as in (\ref{one1}). We denote $S_{i-1}$, $S_{i}$ as independent random variables representing the service times of packets $i-1$ and $i$. Once packet $i-1$ arrives as depicted in Fig. \ref{fig:4} as a large green rectangle, the next inter-arrival time $X_i$ determines the state the next packet will observe, which is determined by the interval the small green rectangle falls in Fig. \ref{fig:4}. $X_i$ also determines how long packet $i$ spends in the system (which is $T_i$). We evaluate $\mathbb{E}[X_iT_i \ | \ (I)]$ in Appendix \ref{app:01}. In the following, we evaluate $\mathbb{E}[X_iT_i \ | \ (s)]$ for $s=W$.

\subsubsection{$\mathbb{E}[X_iT_i \ | \ (W)]$}

Since waiting time is deterministic and the arrivals are Poisson, any packet that arrives in (W) state of the system could arrive at any point in the deterministic interval $[0,\epsilon_I]$ with uniform probability. Hence, the residual waiting time in this case is uniformly distributed: $R_{i-1}^{W} \sim U[0,\epsilon_{I}]$. By replacing the initial deterministic waiting time $\epsilon_{I}$ in the calculation of $\mathbb{E}[X_i T_i \ | \ (I)]$ with uniformly distributed $R_{i-1}^{W}$, we can determine expressions for $\mathbb{E}[X_i T_i \ | \ (W)]$. To this end, we define the function $g_{1/1}(r)$ in Appendix \ref{app:02}. $g_{1/1}(r)$ defines $\mathbb{E}[X_iT_i | R^{W}_{i-1}=r]$ where $R^{W}_{i-1}$ denotes residual time for packet $i-1$ to start service. In particular, we have $g_{1/1}(\epsilon_{I})=\mathbb{E}[X_i T_i \ | \ (I)]$. With this definition, we can express the desired expectation as
\begin{align}\label{eq:exp01}
\mathbb{E}[X_i T_i \ | \ (W)]=\frac{1}{\epsilon_{I}}\int_0^{\epsilon_{I}}g_{1/1}(r)dr
\end{align} 
We obtain closed form expressions for the integral in the RHS of (\ref{eq:exp01}) in Appendix \ref{app:02}.

\subsubsection{$\mathbb{E}[X_iT_i \ | \ (B)]$}

In this case, we first note the following lemma.
\begin{Lemma}
\label{lm:residual}
The residual service time for a packet arriving to the queue in (B) state has the following density function:
\begin{align}
f_R(r)=\frac{\mathbb{P}[S>r]}{\mathbb{E}[S]}
\end{align}
where $S$ represents the general service time. Additionally,
\begin{align}
MGF^{(R)}_{\gamma} = \frac{1-MGF^{(S)}_{\gamma} }{\gamma \mathbb{E}[S]}
\end{align}
\end{Lemma}
In particular, $\mathbb{E}[R|(B)]=\frac{\mathbb{E}[S^2]}{2\mathbb{E}[S]}$. This lemma can be proved by using PASTA property; see also \cite[Eq. (36)]{inoue2018general}. We use the same notation as in (\ref{eq:def}) to define the first and second derivative as $MGF_{\gamma}^{(R,1)}$ and $MGF_{\gamma}^{(R,2)}$. We use $R_{i-1}$ to denote the residual time of packet $i-1$ if it arrives at system state (B) and separately consider the cases $R_{i-1}<X_i$ and $R_{i-1} \geq X_i$, to evaluate the desired expectation $\mathbb{E}[X_i T_i \ | \ (B)]$.

We have $\mbox{Pr}[R_{i-1}<X_i]=MGF^{(R)}_{\lambda}$. Once $R_{i-1}<X_i$, $X_i - r_{i-1}$ is distributed exponentially with the same rate $\lambda$ conditioned on $R_{i-1}=r_{i-1}$. Decomposing $X_i$ as $X_i - R_{i-1}$ plus $R_{i-1}$, we have
\begin{align}\label{eq:exp}
\mathbb{E}[(X_i-R_{i-1}) T_i | (B), E_4]\mbox{Pr}[E_4]&=\mathbb{E}[\tilde{X}_i (\epsilon_I + S_i)]MGF^{(R)}_{\lambda}=\frac{1}{\lambda}(\epsilon_I + \mathbb{E}[S]) MGF^{(R)}_{\lambda} \\
\mathbb{E}[R_{i-1} T_i | (B), E_4] \mbox{Pr}[E_4] &= \mathbb{E}[R_{i-1} | E_4] \mbox{Pr}[E_4](\epsilon_I + \mathbb{E}[S]) = MGF_{\lambda}^{(R,1)}(\epsilon_I + \mathbb{E}[S])
\end{align} 
where the event $E_4$ denotes $R_{i-1} < X_i$ and $\tilde{X}_i$ is an independent exponential random variable with mean $\frac{1}{\lambda}$. Also let $E_5$ be the complement of $E_4$. We evaluate $\mathbb{E}[X_i T_i \ | \ (B), E_5]$ in Appendix \ref{app:03}. We then use the law of total expectation to obtain $\mathbb{E}[X_i T_i \ | \ (B)]=\sum_{i=4}^5 \mathbb{E}[X_iT_i \ | \ (B), E_i]\mbox{Pr}[E_i]$. 
Finally, we have \[\mathbb{E}[X_i T_i]=\sum_{s \in \mathcal{S}_{M/GI/1/1}} \mathbb{E}[X_i T_i  \ | \ (s)]p_s\] and $\mathbb{E}[\Delta]=\lambda \mathbb{E}[X_i T_i] + \frac{1}{\lambda}$.

\subsection{Average Peak AoI}

In this subsection, we will derive expressions for the expected value of PAoI, which is $X_{i^*}+T_{i^{*}}$ where $i^*$ represents smallest packet index among all these served in a service period. We evaluate $\mathbb{E}[X_{i^*} + T_{i^*}]$ as follows: We first calculate $\mathbb{E}[(X_{i}+T_{i})\mathbbm{1}_{i=i^*} \ | \ (s)]$ for all $s \in \mathcal{S}_{M/GI/1/1}$ where $\mathbbm{1}_{i=i^*}$ is the indicator function for the event $i=i^*$. Here, conditioning is on the system state observed by packet $i-1$, denoted as $P_{i-1}$. $\mbox{Pr}[P_{i-1}=(s)]=p_{s}$ where $p_s$, $s \in \mathcal{S}_{M/GI/1/1}$ are as in (\ref{one1}). Note that for an arbitrary arriving packet $i-1$, the next packet index $i$ may or may not be the minimum index in the ensuing service period and the indicator function $\mathbbm{1}_{i=i^*}$ accounts for distinguishing it. Due to ergodicity, the count of favorable cases is performed by conditioning on $P_{i-1}$ and summing over all $s \in \mathcal{S}_{M/GI/1/1}$ yields $\mathbb{E}[(X_{i}+T_{i})\mathbbm{1}_{i=i^*}]$ and we have $\mathbb{E}[X_{i^*}+T_{i^*}]=\frac{\mathbb{E}[(X_{i}+T_{i})\mathbbm{1}_{i=i^*}]}{\mbox{Pr}(i=i^*)}$ where $\mbox{Pr}(i=i^*)$ refers to the probability that an arriving packet $i$ has the minimum index among all those that are served in one service period. We evaluate $\mbox{Pr}(i=i^* \ | \ (s))$ for all $s \in  \mathcal{S}_{M/GI/1/1}$ as well and use total expectation to get $\mbox{Pr}(i=i^*)$. 

We start by noting that conditioned on $P_{i-1}=(B)$, $\mbox{Pr}(i=i^* \ | \ (B))=0$ due to the fact that any packet that arrives in a busy period is served together with the next arriving packet and hence $i$ cannot be the minimum index in this case. We next evaluate $\mathbb{E}[(X_{i}+T_{i})\mathbbm{1}_{i=i^*}|(I)]$ using the format and expressions in Appendix \ref{app:01}.

\subsubsection{$\mathbb{E}[(X_{i}+T_{i})\mathbbm{1}_{i=i^*} \ | \ (I)]$ \text{and} $\mbox{Pr}(i=i^* \ | \ (I))$}

Conditioned on $P_{i-1}=(I)$, among three cases listed in Appendix \ref{app:01} the case of $\epsilon_{I} > X_i$ is ruled out since index $i$ does not yield the maximum $X_j+T_j$ for its service period. On the other hand, this holds for the other two cases. When $\epsilon_{I} - X_i \leq 0, \epsilon_{I} - X_i + S_{i-1} > 0$, $T_i = \epsilon_{I}-X_i+S_{i-1}+\tilde{X}_i+\epsilon_{I}+S_i$. Here $\tilde{X}_i$ is an independent exponentially distributed random variable with mean $\frac{1}{\lambda}$ which represents the arrival time of the packet that comes after the service is finished. Let the event $E_{P1}$ denote $\epsilon_{I} - X_i \leq 0, \epsilon_{I} - X_i + S_{i-1} > 0$. We have
\begin{align}\nonumber
\mathbb{E}[(X_{i}+T_{i})\mathbbm{1}_{i=i^*} \ | \ (I), E_{P1}]\mbox{Pr}[E_{P1}] &= \int_{0}^{\infty} \int^{\epsilon_I + s}_{\epsilon_I} \mathbb{E}[\tilde{X}_i+S_i + 2\epsilon_{I} + s] \lambda e^{-\lambda \hat{x}} f_S(s) d\hat{x} ds \\ &= e^{-\lambda \epsilon_I}\int_{0}^{\infty} (\frac{1}{\lambda} + \mathbb{E}[S]  + 2\epsilon_{I} + s) (1-e^{-\lambda s}) f_S(s) d\hat{x} \nonumber \\ \nonumber &=  e^{-\lambda \epsilon_I} (\frac{1}{\lambda} + 2\mathbb{E}[S]  + 2\epsilon_{I}) \\ &\quad  - e^{-\lambda \epsilon_I}(MGF_{\lambda}^{(S)}(\frac{1}{\lambda} + \mathbb{E}[S]  + 2\epsilon_{I}) + MGF_{\lambda}^{(S,1)}) 
\end{align}
When $\epsilon_{I} + S_{i-1} \leq X_i$, $T_i = \epsilon_{I}+S_{i}$. Let the event $E_{P2}$ denote $\epsilon_{I} + S_{i-1} \leq X_i$. We have
\begin{align}\nonumber
\mathbb{E}[(X_{i}+T_{i})\mathbbm{1}_{i=i^*} \ | \ (I), E_{P2}]\mbox{Pr}[E_{P2}] &= \int_{0}^{\infty} \int_{\epsilon_I+s}^{\infty}\mathbb{E}[S_i + \hat{x} +\epsilon_{I}] \lambda e^{-\lambda \hat{x}} f_S(s) d\hat{x} ds \\ &= e^{-\lambda \epsilon_{I}}\left((\mathbb{E}[S] +2\epsilon_{I}+\frac{1}{\lambda})MGF_{\lambda}^{(S)}+MGF_{\lambda}^{(S,1)}\right)
\end{align}
We finally sum the expressions to get $\mathbb{E}[(X_{i}+T_{i})\mathbbm{1}_{i=i^*}  \ | \ (I)]$:
\begin{align}
\mathbb{E}[(X_{i}+T_{i})\mathbbm{1}_{i=i^*} \ | \ (I)]&=\sum_{k=1}^2\mathbb{E}[(X_{i} + T_{i})\mathbbm{1}_{i=i^*} \ | \ (I),E_{Pk}]\mbox{Pr}[E_{Pk}] \\ &= e^{-\lambda \epsilon_I} (\frac{1}{\lambda} + 2\mathbb{E}[S]  + 2\epsilon_{I}) 
\end{align}
We calculate $\mbox{Pr}(i=i^* \ | \ (I))$ as follows:
\begin{align}
\mbox{Pr}(i=i^* \ | \ (I)) = 1-\mbox{Pr}(\epsilon_{I} > X_i) = e^{-\lambda \epsilon_I}
\end{align}

\subsubsection{$\mathbb{E}[(X_{i}+T_{i})\mathbbm{1}_{i=i^*} \ | \ (W)]$ \text{and} $\mbox{Pr}(i=i^* \ | \ (W))$}

Conditioned on $P_{i-1}=(W)$, the residual waiting time in this case is uniformly distributed: $R_{i-1}^{W} \sim U[0,\epsilon_{I}]$. We let $R_{i-1}^{W}=r$ and define $g^{(P)}_{1/1}(r)=\mathbb{E}[(X_{i}+T_{i})\mathbbm{1}_{i=i^*} \ | \ (W), R_{i-1}^{W}=r]$ as
\begin{align}\nonumber
g^{(P)}_{1/1}(r)\triangleq  e^{-\lambda r} (\frac{1}{\lambda} + 2\mathbb{E}[S]  + \epsilon_{I} + r) 
\end{align}
We have $g^{(P)}_{1/1}(\epsilon_I)=\mathbb{E}[(X_{i}+T_{i})\mathbbm{1}_{i=i^*} \ | \ (I)]$ and additionally
\begin{align}\label{eq:p1}
\mathbb{E}[(X_{i}+T_{i})\mathbbm{1}_{i=i^*} \ | \ (W)]&=\frac{1}{\epsilon_{I}}\int_0^{\epsilon_{I}}g^{(P)}_{1/1}(r)dr \\&= \frac{1}{\epsilon_I} \left((\frac{1}{\lambda} + 2\mathbb{E}[S]  + \epsilon_{I})\frac{1}{\lambda}(1-e^{-\lambda \epsilon_I}) + \frac{1}{\lambda^2}(1-e^{-\lambda \epsilon_I}(1+\lambda \epsilon_I))  \right) 
\end{align} 
We calculate $\mbox{Pr}(i=i^* \ | \ (W))$ as follows:
\begin{align}
\mbox{Pr}(i=i^* \ | \ (W)) = \frac{1}{\epsilon_I} \int_0^{\epsilon_I} e^{-\lambda r} dr =\frac{1 - e^{-\lambda \epsilon_I}}{\lambda \epsilon_I}
\end{align}
Finally, we have \[\mathbb{E}[(X_{i}+T_{i})\mathbbm{1}_{i=i^*}]=\sum_{s \in \mathcal{S}_{M/GI/1/1}} \mathbb{E}[(X_{i} + T_{i})\mathbbm{1}_{i=i^*} \ | \ (s)]p_s\] and $\mathbb{E}[X_{i^*}+T_{i^*}] = \frac{\mathbb{E}[(X_{i}+T_{i})\mathbbm{1}_{i=i^*}]}{\mbox{Pr}(i=i^*)}$. Here, we calculate $\mbox{Pr}(i=i^*)$ as 
 \[ \mbox{Pr}(i=i^*) = p_I\mbox{Pr}(i=i^* \ | \ (I)) + p_W\mbox{Pr}(i=i^* \ | \ (W)) =  p_Ie^{-\lambda \epsilon_I} + p_W \frac{1 - e^{-\lambda \epsilon_I}}{\lambda \epsilon_I} \]
In compact form, we get $\mbox{Pr}(i=i^*) = \frac{1}{\lambda T_{cycle}}$ and
\begin{align}\label{eq:compact}
\mathbb{E}[X_{i^*}+T_{i^*}]= \epsilon_I - \frac{1}{\lambda} e^{-\lambda \epsilon_I} + \frac{2}{\lambda} + 2\mathbb{E}[S]
\end{align}
We observe that $\mathbb{E}[X_{i^*}+T_{i^*}]$ is monotone increasing with $\epsilon_I$.

\section{Average AoI and Average Peak AoI for M/GI/1/$2^*$ with Waiting}
\label{sec:eval}

In this section, we consider average AoI and average peak AoI for M/GI/1/$2^*$ packet management with waiting.

\subsection{Average AoI}

To calculate $\mathbb{E}[\Delta]$ in (\ref{aaoi}), we will find the correlation between $X$ and $T$. Recall that $X$ has marginal exponential distribution with rate $\lambda$. The system can be in four different states. In view of the renewal structure, we have the following stationary probabilities for each state:
\begin{align}\label{one}
p_{I} &= \frac{1}{\lambda T_{cycle}}, \ p_{B}= \frac{\mathbb{E}[S]}{T_{cycle} MGF^{(S)}_{\lambda}} \\
p_{WaB} & = \frac{\epsilon_{B}}{T_{cycle}}\left(\frac{1}{MGF^{(S)}_{\lambda}}-1\right), \ p_{WaI}= \frac{\epsilon_{I}}{T_{cycle}} \label{two}
\end{align}
where $T_{cycle}$ is the expected length of one renewal cycle:
\begin{align}\hspace{-0.1in}
T_{cycle}=\frac{1}{\lambda}+\epsilon_{I}+\epsilon_{B}(\frac{1}{MGF^{(S)}_{\lambda}}-1)+\frac{\mathbb{E}[S]}{MGF^{(S)}_{\lambda}}
\end{align}
Note that in one renewal cycle in M/GI/1/$2^*$ queuing model, the system first starts in (I) state and shifts to (WaI) when an arrival occurs. The expected length of staying in (I) is $\frac{1}{\lambda}$. After staying in (WaI) for a deterministic $\epsilon_I$ time units, the system state switches to (B) and stays there for a service time. The system may repeatedly switch between (WaB) and (B) in a single cycle. If an arrival occurs during service time, the system state switches to (WaB) and after a deterministic $\epsilon_B$ units, the system state goes back to (B). If no arrival occurs in one service time, then system state shifts to (I) and this completes one renewal cycle. The expected length of back and forth between (B) and (WaB) is: 
\begin{align*}
 \mathbb{E}[\sum_{n=1}^N S_n + \sum_{n=2}^N \epsilon_B]  
 \end{align*}
where $S_n$ represents independent realizations of i.i.d. service time variables and $N$ is a geometric random variable with stop probability $\mbox{Pr}[X>S]=MGF_{\lambda}^{(S)}$. We can then evaluate $\mathbb{E}[\sum_{n=1}^N S_n]=\frac{\mathbb{E}[S]}{MGF_{\lambda}^{(S)}}$ by Wald's identity \cite{gallager2012discrete} and $\mathbb{E}[\sum_{n=2}^N \epsilon_B]=\epsilon_B (\frac{1}{MGF_{\lambda}^{(S)}}-1)$. By Renewal Reward Theorem, the stationary probabilities $p_s$, $s \in \mathcal{S}_{M/GI/1/2*}=\{\mbox{I},\mbox{WaI},\mbox{B},\mbox{WaB}\}$, in (\ref{one})-(\ref{two}) are equal to the expected time spent in system state $s$ in one renewal cycle divided by the expected cycle length.

\begin{figure}[!t]
\centering{
\hspace{-0.3cm} 
\includegraphics[totalheight=0.25\textheight]{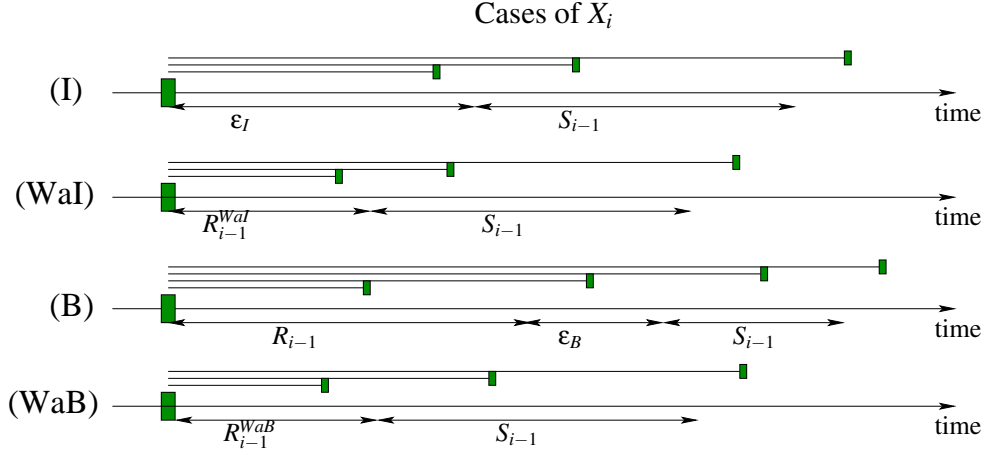}}\vspace{-0.1in}
\caption{\sl Four states of the system packet $i-1$ can observe in M/GI/1/$2^*$ scheme.}
\label{fig:2} 
\vspace{-0.2in}
\end{figure}

We next evaluate $\mathbb{E}[X_i T_i \ | \ (s)]$ for $s \in \mathcal{S}_{M/GI/1/2^*}$ and conditioning is on the system state observed by packet $i-1$, denoted as $P_{i-1}$. Due to PASTA property, $\mbox{Pr}[P_{i-1}=(s)]=p_{s}$ where $p_s$, $s \in \mathcal{S}_{M/GI/1/2^*}$ are as in (\ref{one})-(\ref{two}). We denote $S_{i-1}$, $S_{i}$ as independent random variables representing the service times of packets $i-1$ and $i$. Once packet $i-1$ arrives (shown in Fig. \ref{fig:2} as a big green rectangle), the next inter-arrival time $X_i$ determines the next state (the interval the small green rectangle falls in Fig. \ref{fig:2}) and how long packet $i$ spends in the system (which is $T_i$). We defer the evaluation of $\mathbb{E}[X_i T_i \ | \ (I)]$ to Appendix \ref{app:1} and start with conditioning on (WaI).

\subsubsection{$\mathbb{E}[X_i T_i \ | \ \mbox{(WaI)}]$}

Since waiting time is deterministic and the arrivals are Poisson, any packet that arrives in (WaI) state of the system could arrive at any point in this deterministic interval with uniform probability. Hence, the residual waiting time in this case is uniformly distributed: $R_{i-1}^{WaI} \sim U[0,\epsilon_{I}]$. By replacing the initial deterministic waiting time $\epsilon_{I}$ in the calculation of $\mathbb{E}[X_i T_i \ | \ (I)]$ with uniformly distributed $R_{i-1}^{WaI}$, we can determine expressions for $\mathbb{E}[X_i T_i \ | \ (WaI)]$. To this end, we define the function $g_{1/2^*}(r)$ defines $\mathbb{E}[X_iT_i \ | \ R^{WaI}_{i-1}=r]$ in Appendix \ref{app:2}. In particular, we have $g_{1/2^*}(\epsilon_{I})=\mathbb{E}[X_i T_i \ | \ (I)]$. With this definition, we express the desired expectation as
\begin{align}\label{eq:exp1}
\mathbb{E}[X_i T_i \ | \ (WaI)]=\frac{1}{\epsilon_{I}}\int_0^{\epsilon_{I}}g_{1/2^*}(r)dr
\end{align} 
We obtain closed form expressions for the integral in the RHS of (\ref{eq:exp1}) in Appendix \ref{app:4}.

\subsubsection{$\mathbb{E}[X_i T_i \ | \ \mbox{(WaB)}]$}

Due to identical reasoning to the (WaI) case, the residual waiting time in (WaB) state is uniformly distributed: $R_{i-1}^{WaB} \sim U[0,\epsilon_{B}]$. We thus have the following expression for the expectation: 
\begin{align}\label{eq:exp2}
\mathbb{E}[X_i T_i \ | \ (WaB)]=\frac{1}{\epsilon_{B}}\int_0^{\epsilon_{B}}g_{1/2^*}(r)dr
\end{align} 
Closed form expressions for (\ref{eq:exp2}) are obtained in Appendix \ref{app:4}.

\subsubsection{$\mathbb{E}[X_i T_i \ | \ \mbox{(B)}]$}

Residual time in this case is given in Lemma \ref{lm:residual}. It is remarkable that residual time for a packet arriving in busy state is invariant with respect to the packet management scheme. We have $\mbox{Pr}[R_{i-1}<X_i]=MGF^{(R)}_{\lambda}$. When $R_{i-1}<X_i$, in this case $X_i - r_{i-1}$ is distributed exponentially with the same rate $\lambda$ conditioned on $R_{i-1}=r_{i-1}$. Hence, we get the same form of expressions by replacing $X_i - r_{i-1}$ with $X_i$ and $\epsilon_{I}$ with $\epsilon_{B}$. Decomposing $X_i$ as $X_i - R_{i-1}$ plus $R_{i-1}$, we have
\begin{align}\label{eq:exp}
\mathbb{E}[(X_i-R_{i-1}) T_i \ | \ (B), E_4]=g_{1/2^*}(\epsilon_{B})
\end{align} 
where the event $E_4$ denotes $R_{i-1} < X_i$. Also let $E_5$ be the complement of $E_4$. We evaluate $\mathbb{E}[R_{i-1} T_i \ | \ (B), E_4]$ and $\mathbb{E}[X_i T_i \ | \ (B), E_5]$ in Appendix \ref{app:3}.

Having evaluated all cases, we can now use the law of total expectation to obtain \[ \mathbb{E}[X_i T_i | (B)]=\sum_{i=4}^5 \mathbb{E}[X_iT_i \ | \ (B), E_i]\mbox{Pr}[E_i] \] Finally, we have \[\mathbb{E}[X_i T_i]=\sum_{s \in \mathcal{S}_{M/GI/1/2^*}} \mathbb{E}[X_i T_i \ | \ (s)]p_s\] and $\mathbb{E}[\Delta]=\lambda \mathbb{E}[X_i T_i] + \frac{1}{\lambda}$.

\subsection{Average Peak AoI}

In this subsection, we focus on average PAoI by following the path we followed earlier for M/GI/1/1 scheme through evaluating $\mbox{E}[(X_{i}+T_{i})\mathbbm{1}_{i=i^*} \ | \ (s)]$ and $\mbox{Pr}(i=i^* \ | \ (s))$ for all $s \in \mathcal{S}_{M/GI/1/2^*}$. For an arbitrary arriving packet $i-1$, the next packet index $i$ may or may not be the minimum index in its service period. We next evaluate $\mbox{E}[(X_{i}+T_{i})\mathbbm{1}_{i=i^*} \ | \ (I)]$ using the format and expressions in Appendix \ref{app:1}.

\subsubsection{$\mathbb{E}[(X_{i}+T_{i})\mathbbm{1}_{i=i^*} \ | \ (I)]$ \text{and} $\mbox{Pr}(i=i^* \ | \ (I))$}

Conditioned on $P_{i-1}=(I)$, among three cases listed in Appendix \ref{app:1} the case of $\epsilon_{I} > X_i$ is ruled out since index $i$ does not yield the maximum $X_j+T_j$ for its service period. On the other hand, this holds for the other two cases. When $\epsilon_{I} - X_i \leq 0, \epsilon_{I} - X_i + S_{i-1} > 0$, $T_i = \epsilon_{I}-X_i+S_{i-1}+\epsilon_{B}+S_i$. Let the event $\widehat{E}_{P1}$ denote $\epsilon_{I} - X_i \leq 0, \epsilon_{I} - X_i + S_{i-1} > 0$. We have
\begin{align}\nonumber
\mathbb{E}[(X_{i}+T_{i})\mathbbm{1}_{i=i^*} \ | \ (I), \widehat{E}_{P1}]\mbox{Pr}[\widehat{E}_{P1}] &= \int_{0}^{\infty} \int^{\epsilon_I + s}_{\epsilon_I} \mathbb{E}[S_i + \epsilon_{I} + \epsilon_B + s] \lambda e^{-\lambda \hat{x}} f_S(s) d\hat{x} ds \\ &= e^{-\lambda \epsilon_I}\int_{0}^{\infty} (\mathbb{E}[S]  + \epsilon_{I} + \epsilon_B + s) (1-e^{-\lambda s}) f_S(s) d\hat{x} \nonumber \\ \nonumber &=  e^{-\lambda \epsilon_I} (2\mathbb{E}[S]  + \epsilon_{I} + \epsilon_B) \\ &\quad  - e^{-\lambda \epsilon_I}(MGF_{\lambda}^{(S)}(\mathbb{E}[S]  + \epsilon_{I} + \epsilon_B) + MGF_{\lambda}^{(S,1)}) 
\end{align}
When $\epsilon_{I} + S_{i-1} \leq X_i$, $T_i = \epsilon_{I}+S_{i}$. Let the event $\widehat{E}_{P2}$ denote $\epsilon_{I} + S_{i-1} \leq X_i$. We have
\begin{align}\nonumber
\mathbb{E}[(X_{i}+T_{i}) \mathbbm{1}_{i=i^*} \ | \ (I), \widehat{E}_{P2}]\mbox{Pr}[\widehat{E}_{P2}] &= \int_{0}^{\infty} \int_{\epsilon_I+s}^{\infty}\mathbb{E}[S_i + \hat{x} +\epsilon_{I}] \lambda e^{-\lambda \hat{x}} f_S(s) d\hat{x} ds \\ &= e^{-\lambda \epsilon_{I}}\left((\mathbb{E}[S] +2\epsilon_{I}+\frac{1}{\lambda})MGF_{\lambda}^{(S)}+MGF_{\lambda}^{(S,1)}\right) 
\end{align}
We finally sum the expressions to get $\mathbb{E}[(X_{i} + T_{i})\mathbbm{1}_{i=i^*} \ | \ (I)]$:
\begin{align}
\mathbb{E}[(X_{i} + T_{i})\mathbbm{1}_{i=i^*} \ | \ (I)]&=\sum_{k=1}^2\mathbb{E}[(X_{i} + T_{i})\mathbbm{1}_{i=i^*} | (I),\widehat{E}_{Pk}]\mbox{Pr}[\widehat{E}_{Pk}] \\ &= e^{-\lambda \epsilon_I} ( 2\mathbb{E}[S]  + \epsilon_{I} + \epsilon_B  + MGF_{\lambda}^{(S)}(\frac{1}{\lambda}  + \epsilon_{I} - \epsilon_B)) 
\end{align}
In this case, we calculate $\mbox{Pr}(i=i^* \ | \ (I))$ as
\begin{align}
\mbox{Pr}(i=i^* \ | \ (I)) = 1- \mbox{Pr}(\epsilon_I > X_i) = e^{-\lambda \epsilon_I}
\end{align}

\subsubsection{$\mathbb{E}[ (X_{i}+T_{i})\mathbbm{1}_{i=i^*} \ | \ (WaI)]$ \text{and} $\mbox{Pr}(i=i^* \ | \ (WaI))$}

Conditioned on $P_{i-1}=(WaI)$, the residual waiting time in this case is uniformly distributed: $R_{i-1}^{WaI} \sim U[0,\epsilon_{I}]$. We let $R_{i-1}^{WaI}=r$ and define $g^{(P)}_{1/2^*}(r) = \mathbb{E}[ (X_{i}+T_{i})\mathbbm{1}_{i=i^*} \ | \ (WaI), R_{i-1}^{(WaI)}=r]$ as 
\begin{align}\nonumber
g^{(P)}_{1/2^*}(r)\triangleq e^{-\lambda r} (2\mathbb{E}[S]  + r + \epsilon_B  + MGF_{\lambda}^{(S)}(\frac{1}{\lambda}  + \epsilon_I - \epsilon_B)) \end{align}
We have $g^{(P)}_{1/2^*}(\epsilon_I)=\mathbb{E}[(X_{i} + T_{i})\mathbbm{1}_{i=i^*} \ | \ (I)]$ and additionally
\begin{align}\label{eq:ppk1}
\mathbb{E}[(X_{i} + T_{i})\mathbbm{1}_{i=i^*} | (WaI)]&=\frac{1}{\epsilon_{I}}\int_0^{\epsilon_{I}}g^{(P)}_{1/2^*}(r)dr \\&= \frac{1}{\epsilon_I} \Big((2\mathbb{E}[S]  + \epsilon_{B} + MGF_{\lambda}^{(S)}(\frac{1}{\lambda} + \epsilon_I  - \epsilon_B))\frac{1}{\lambda}(1-e^{-\lambda \epsilon_I}) \nonumber \\ &\qquad + \frac{1}{\lambda^2}(1-e^{-\lambda \epsilon_I}(1+\lambda \epsilon_I))  \Big) 
\end{align} 
We calculate $\mbox{Pr}(i=i^* \ | \ (WaI))$ as follows:
\begin{align}
\mbox{Pr}(i=i^* \ | \ (WaI)) &= \frac{1}{\epsilon_I} \int_0^{\epsilon_I} e^{-\lambda r} dr \\ &=\frac{1 - e^{-\lambda \epsilon_I}}{\lambda \epsilon_I}
\end{align}

\subsubsection{$\mathbb{E}[ (X_{i}+T_{i})\mathbbm{1}_{i=i^*} \ | \ (WaB)]$ \text{and} $\mbox{Pr}(i=i^* \ | \ (WaB))$}

Conditioned on $P_{i-1}=(WaB)$, the residual waiting time in this case is uniformly distributed: $R_{i-1}^{WaB} \sim U[0,\epsilon_{B}]$. We have
\begin{align}\label{eq:exppk2}
\mathbb{E}[(X_{i} + T_{i})\mathbbm{1}_{i=i^*} | (WaB)]&=\frac{1}{\epsilon_{B}}\int_0^{\epsilon_{B}}g^{(P)}_{1/2^*}(r)dr  \\&= \frac{1}{\epsilon_B} \Big((2\mathbb{E}[S]  + \epsilon_{B} + MGF_{\lambda}^{(S)}(\frac{1}{\lambda}  + \epsilon_I - \epsilon_B))\frac{1}{\lambda}(1-e^{-\lambda \epsilon_B}) \nonumber \\ &\qquad + \frac{1}{\lambda^2}(1-e^{-\lambda \epsilon_B}(1+\lambda \epsilon_B))  \Big) 
\end{align} 
We calculate $\mbox{Pr}(i=i^* \ | \ (WaB))$ as follows:
\begin{align}
\mbox{Pr}(i=i^* \ | \ (WaB)) &= \frac{1}{\epsilon_B} \int_0^{\epsilon_B} e^{-\lambda r} dr \\ &=\frac{1 - e^{-\lambda \epsilon_B}}{\lambda \epsilon_B}
\end{align}

\subsubsection{$\mathbb{E}[ (X_{i}+T_{i})\mathbbm{1}_{i=i^*} \ | \ (B)]$ \text{and} $\mbox{Pr}(i=i^* \ | \ (B))$}

Conditioned on $P_{i-1}=(B)$, there are two cases to consider: $R_{i-1}<X_i$ denoted as the event $\widehat{E}_{P3}$ and $R_{i-1}\geq X_i$ denoted as the event $\widehat{E}_{P4}$. We have $\mbox{Pr}(i=i^* \ | \ (B), \widehat{E}_{P4})=0$ due to the fact that any packet $i$ that arrives in a busy period after the packet $i-1$ arriving in the same busy period is served together with it and thus $i$ cannot be the minimum index in this case. Hence, we rule out $\widehat{E}_{P4}$. We have $\mbox{Pr}[\widehat{E}_{P3}]=MGF^{(R)}_{\lambda}$. In the event $\widehat{E}_{P3}$, $X_i - R_{i-1} \sim \tilde{X}_i$ is distributed exponentially with the same rate $\lambda$. The system has the same state as (I) after $R_{i-1}$ waiting with the exception that initial $\epsilon_I$ waiting is replaced with $\epsilon_B$. We have
\begin{align}\nonumber
\hspace{-0.1in}\mathbb{E}[(X_{i}+T_{i})\mathbbm{1}_{i=i^*} | (B)] &= \mathbb{E}[(X_{i} - R_{i-1}+T_{i})\mathbbm{1}_{i=i^*} | (B), E_{P3}]\mbox{Pr}[E_{P3}] \\ \label{eq:expp} &\qquad \qquad \qquad+ \mathbb{E}[R_{i-1}\mathbbm{1}_{i=i^*}  | (B), E_{P3}]\mbox{Pr}[E_{P3}] \\ &= g_{1/2^*}^{(P)}(\epsilon_B)MGF^{(R)}_{\lambda} + e^{-\lambda \epsilon_B}MGF^{(R,1)}_{\lambda} 
\end{align} 
We calculate $\mbox{Pr}(i=i^* \ | \ (B))$ as follows:
\begin{align}
\mbox{Pr}(i=i^* \ | \ (B)) = e^{-\lambda \epsilon_B} MGF_{\lambda}^{(R)} 
\end{align}
We finally combine our findings to get \[\mathbb{E}[(X_{i} + T_{i})\mathbbm{1}_{i=i^*}]=\sum_{s \in \mathcal{S}_{M/GI/1/2^*}} \mathbb{E}[(X_{i} + T_{i})\mathbbm{1}_{i=i^*}  \ | \ (s)]p_s\] and $\mathbb{E}[X_{i^*}+T_{i^*}] = \frac{\mathbb{E}[(X_{i}+T_{i})\mathbbm{1}_{i=i^*}]}{\mbox{Pr}(i=i^*)}$. Here, we calculate $\mbox{Pr}(i=i^*)$ through the summation $\mbox{Pr}(i=i^*) = \sum_{s \in \mathcal{S}_{M/GI/1/2^*}} p_s\mbox{Pr}(i=i^* \ | \ (s))$. We work on the expression of average peak AoI to simplify it to the following form:
\begin{align}\nonumber
\mathbb{E}[(X_{i} + T_{i})\mathbbm{1}_{i=i^*}] =  &\epsilon_B (1-MGF_{\lambda}^{(S)}) - MGF_{\lambda}^{(S,1)}e^{-\lambda \epsilon_B} + MGF_{\lambda}^{(S)} \epsilon_I  \\ &- \frac{MGF_{\lambda}^{(S)}}{\lambda} e^{-\lambda \epsilon_I} + 2 \mathbb{E}[S] + \frac{1}{\lambda}(1+MGF_{\lambda}^{(S)})  \label{eq:compact2}
\end{align}
where we calculate additionally that $\mbox{Pr}(i=i^*) =\frac{1}{\lambda MGF_{\lambda}^{(S)} T_{cycle}}$. We observe that the expression in (\ref{eq:compact2}) is monotone increasing with $\epsilon_I$ and $\epsilon_B$ as $MGF_{\lambda}^{(S)}$ and $MGF_{\lambda}^{(S,1)}$ are both nonnegative and $MGF_{\lambda}^{(S)} \leq 1$.

\section{Numerical Results and Discussion}
\label{sec:Numres}

In this section, we provide numerical comparisons for average AoI, average peak AoI and the tradeoff between the two for M/GI/1/1 and M/GI/1/$2^*$ packet management schemes. We start by noting that the average peak AoI is monotonically increasing with waiting for both schemes. We have already observed this in the expression of average peak AoI for M/GI/1/1 scheme in (\ref{eq:compact}) and for M/GI/1/$2^*$ in (\ref{eq:compact2}). Therefore, an improved average AoI enabled by introducing waiting before serving comes at the cost of increased average peak AoI. To understand the tradeoff between average AoI and average peak AoI, we will consider optimizing the waiting period with the objective of weighted sum of AoI and average peak AoI for different weights introduced according to the importance of each. For M/GI/1/1, we consider
\begin{align}
\min_{\epsilon_I \geq 0} \omega_1 \mathbb{E}[\Delta] + \omega_2 \mathbb{E}[PAoI]
\end{align}
For M/GI/1/$2^*$, we consider
\begin{align}
\min_{\epsilon_I \geq 0, \epsilon_B \geq 0} \omega_1 \mathbb{E}[\Delta] + \omega_2 \mathbb{E}[PAoI]
\end{align}
where $\omega_1, \omega_2 \geq 0$ are the weights of average AoI and average peak AoI, respectively. Covering all possible weights enables us to obtain the tradeoff curves between average AoI and average peak AoI. Note that minimizing average peak AoI requires setting the waiting periods to zero whereas this is not the case if the objective is to minimize average AoI. 

In here, we determine optimal deterministic waiting through exhaustive search over all $\epsilon_I$ in M/GI/1/1 and over all $(\epsilon_I,\epsilon_B)$ pairs in M/GI/1/$2^*$. In our numerical experiments, we observe that $\mathbb{E}[\Delta]$ is always quasi-convex with respect to $(\epsilon_I,\epsilon_B)$ and hence we assert existence of an optimal pair $(\epsilon_I,\epsilon_B)$. It is indeed not very hard to show that as $\epsilon_I\rightarrow \infty$ and $\epsilon_B \rightarrow \infty$ individually, $\mathbb{E}[\Delta]$ also grows to $\infty$, guaranteeing a bounded $(\epsilon_I,\epsilon_B)$ that minimizes $\mathbb{E}[\Delta]$. We will explore rigorous details of optimization of the waiting in future work. Ultimately, our reporting as ``optimal" is based on numerical observations given that the function to be optimized is a single variable one in M/GI/1/1 and a two variable one in M/GI/1/$2^*$ and it is available in closed form in both cases.   

We test the tradeoff between average AoI and average peak AoI under Inverse Gaussian and Gamma service distributions. We use the closed-form analytical expressions derived in previous sections to obtain the plots in this section. Additionally, we verified these plots using packet-based simulations using random number generators in MATLAB where we use a minimum of $10^6$ packets (around $10^8$ for longer tailed cases) for convergence. We observe in each case that these expressions are accurate.

\subsection{Inverse Gaussian Service Distribution}

Inverse Gaussian distribution is defined as $f_S(s)=\sqrt{\frac{\alpha}{2\pi s^3}}e^{-\alpha \frac{(s-1/\mu)^2}{2s/\mu^2}}$ for $s \geq 0$. For this distribution, $\mathbb{E}[S]=\frac{1}{\mu}$ and $\alpha$ is the shape parameter that determines the variance and tail behavior. As $\alpha$ gets smaller, the tail gets heavier. We have the following closed-form expressions for Inverse Gaussian distribution:
\begin{align*}
MGF^{(S)}_{\lambda}&=e^{\alpha \mu \left(1-\sqrt{1+2\lambda/(\alpha \mu^2)}\right)} \\
MGF^{(S,1)}_{\lambda}&=\frac{MGF^{(S)}_{\lambda}}{\mu \sqrt{1+2\lambda/(\alpha \mu^2)}} \\
MGF^{(S,2)}_{\lambda}&=\frac{MGF^{(S,1)}_{\lambda}}{\mu\sqrt{1+2\lambda/(\alpha \mu^2)}} + \frac{MGF^{(S)}_{\lambda}}{\alpha \mu^3  \left(1+2\lambda/(\alpha \mu^2)\right)^{\frac{3}{2}}}
\end{align*}

\begin{figure}[t] 
%\vspace{-3mm}
    \centering
\subfigure[]
    {
    \includegraphics[totalheight=0.35\textheight]{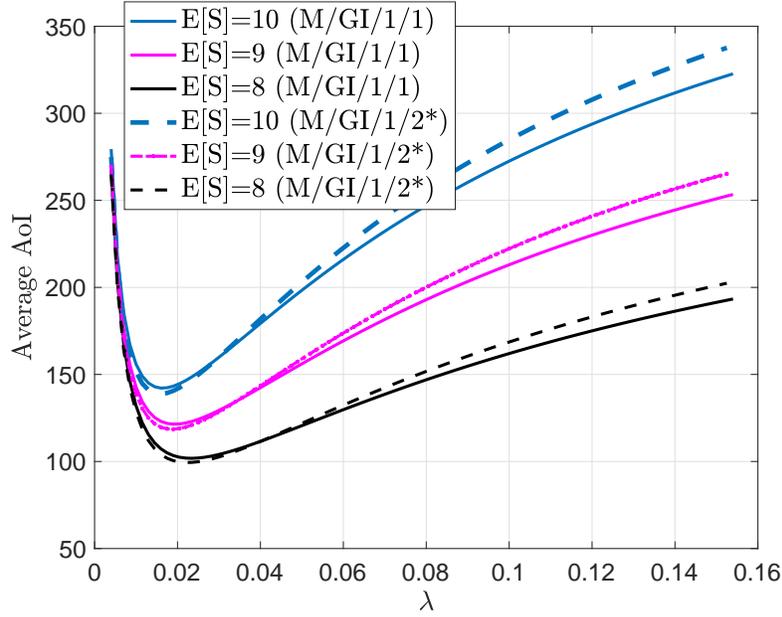}
    \label{fig:local1}} 
\subfigure[]
    {
    \includegraphics[totalheight=0.35\textheight]{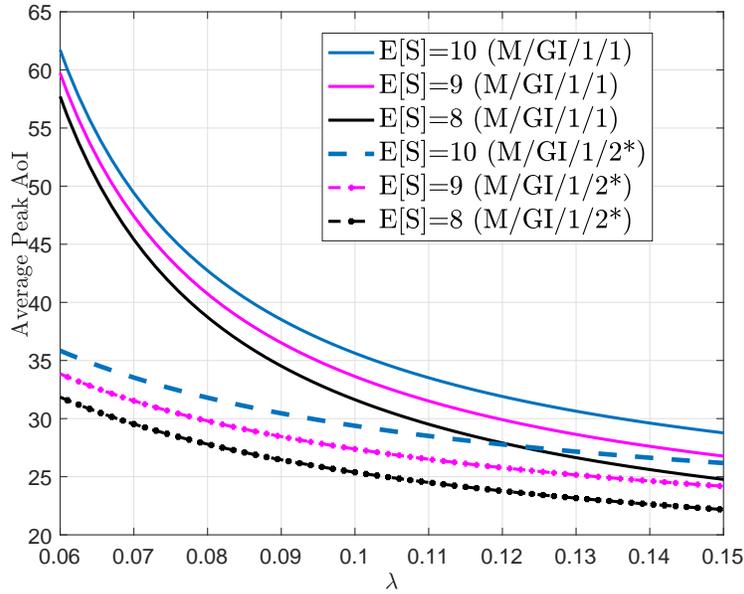}
    \label{fig:local2}} 
    \vspace{-1mm} 
 \caption{\sl The plots are for average AoI and average peak AoI versus $\lambda$ for different mean service rate values under inverse Gaussian service distribution with $\alpha=0.1$ and M/GI/1/1 and M/GI/1/$2^*$ packet management schemes. (a) Average AoI, (b) Average peak AoI. \vspace{-3mm}}
  \label{fig:local001} 
\end{figure}

In Figs. \ref{fig:local1} and \ref{fig:local2}, we plot the average AoI and average peak AoI versus arrival rate $\lambda$ for different mean service rates under zero waiting for the inverse Gaussian service distribution with shape parameter $\alpha=0.1$ for M/GI/1/1 and M/GI/1/$2^*$ schemes. The range of arrival rate $\lambda$ is chosen smaller in Fig. \ref{fig:local2} with respect to that in Fig. \ref{fig:local1} especially to emphasize the order in between different plots. We observe that average AoI monotonically increases with $\mathbb{E}[S]$, and attains its minima at smaller values of $\lambda$ as the service rate becomes smaller. Similarly, we observe that average peak AoI is monotone decreasing with $\lambda$ for all cases considered in this paper. The range of $\lambda$ shown in Fig. \ref{fig:local2} is selected to compare the plots as the increase in average peak AoI is dramatic for all queuing schemes once $\lambda$ drops below shown range. It is also remarkable to see that the order between average AoI and average peak AoI do not match. These curves reveal the cost paid in terms of average peak AoI while minimizing average AoI through introducing waiting before service starts. We also note that average peak AoI is smaller than average AoI. This is counterintuitive in that average peak AoI is expected to be larger than average AoI. However, a closer look at the definitions of average AoI and average peak AoI reveals that average AoI is related to the second order statistic of the service process whereas average peak AoI is related to the first order statistic of the service process. Since in our examples the variance of the service process is selected to be high while the mean is relatively small, the scale of average AoI is affected more dominantly with respect to the average peak AoI. As a consequence, we make the counterintuitive observation that for both packet management schemes average AoI is larger than average peak AoI. 

\begin{figure}[t] 
%\vspace{-3mm}
    \centering
\subfigure[]
    {
    \includegraphics[totalheight=0.35\textheight]{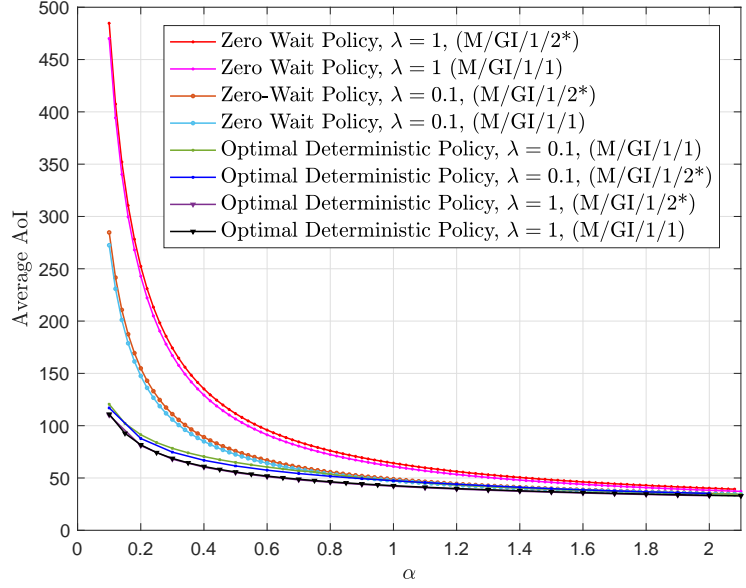}
    \label{fig:local21}} 
\subfigure[]
    {
    \includegraphics[totalheight=0.35\textheight]{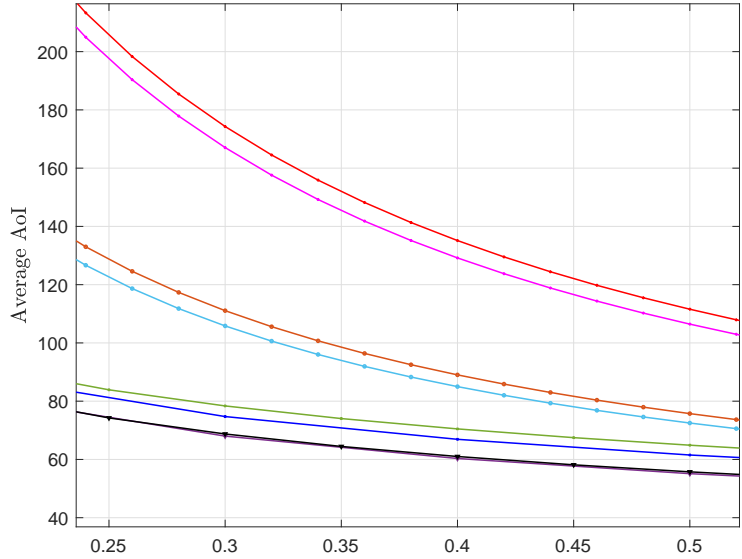}
    \label{fig:local22}} 
    \vspace{-0mm} 
 \caption{\sl The plots are for average AoI versus $\alpha$ under inverse Gaussian service distribution with $\mu=0.1$ and show the comparison of zero waiting versus optimal deterministic waiting under M/GI/1/1 and M/GI/1/$2^*$ schemes. (b) is zoomed version of (a). \vspace{-3mm}}
  \label{fig:local3} 
\end{figure}

In Fig. \ref{fig:local21}, we show average AoI versus $\alpha$ under inverse Gaussian service distribution with $\mu=0.1$ and for $\lambda=0.1$ and $\lambda=1$ for both M/GI/1/1 and M/GI/1/$2^*$ schemes. Fig. \ref{fig:local22} is the zoomed version of Fig. \ref{fig:local21}. Note that the service distribution has a decreasing variance for fixed mean service rate and increasing $\alpha$. In all of our numerical experiments, we observe invariantly that average AoI increases and the improvement brought by waiting is more significant for larger variances with fixed mean. In Figs. \ref{fig:local21} and \ref{fig:local22}, we directly address this point. We provide comparisons of AoI performances for zero-waiting and optimal deterministic waiting in all cases. We observe that the improvement brought by waiting could be as large as 75\% for $\alpha=0.1$ and $\lambda=1$ for both packet management schemes. In view of \cite{sun2017update}, such gains are especially expected at high system loads and long-tailed service distributions as is the case for the particular inverse Gaussian distribution. We also observe that as the system load is decreased, the improvement brought by waiting also decreases; still, it is quite significant (around 60 \%) when $\lambda=0.1$ under heavy tail case $\alpha=0.1$. The zero wait performance of M/GI/1/1 is higher than that for M/GI/1/$2^*$. In contrast, we note that the average AoI under optimal deterministic waiting for M/GI/1/1 and M/GI/1/$2^*$ are almost identical. These comparisons are verified for different parameter values and makes one question the value of a data buffer in the presence of waiting before serving option.

\begin{figure}[!t]
\centering{
%\hspace{-0.3cm} 
\includegraphics[totalheight=0.35\textheight]{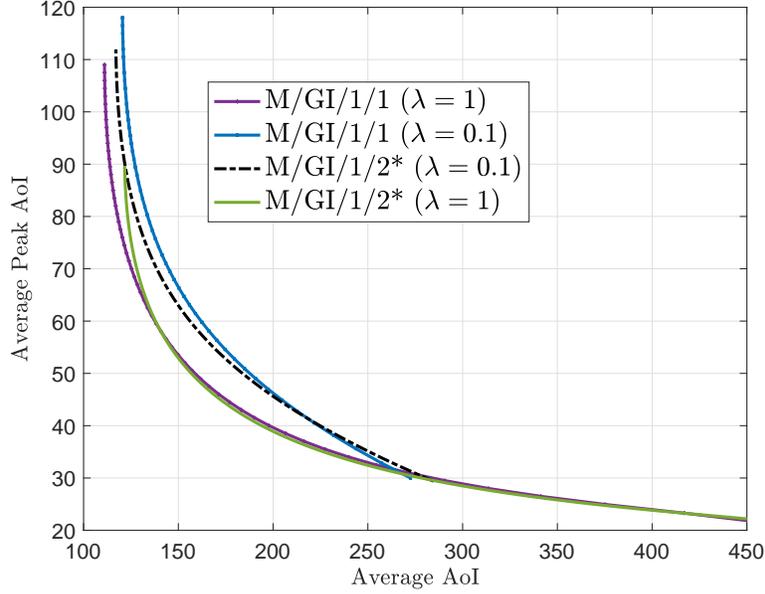}}\vspace{-0.1in}
\caption{\sl The tradeoff curves between average AoI and average peak AoI for M/GI/1/1 and M/GI/1/$2^*$ schemes.}
\label{fig:local3} 
\end{figure}

In Fig. \ref{fig:local3}, we observe the tradeoff curves for achieved average AoI versus achieved average peak AoI under both packet management schemes with different arrival rate $\lambda$. This plot is obtained by varying the waiting period $\epsilon_I$. The service rate is set to $\mu=0.1$ and shape parameter is $\alpha=0.1$. It is remarkable that M/GI/1/1 achieves a better tradeoff with respect to M/GI/1/$2^*$ for $\lambda=1$ whereas this observation is in the other direction for $\lambda=0.1$.

\subsection{Gamma Service Distribution}
Gamma distribution is defined as $f_S(s)=\frac{k^k \mu^k}{\Gamma(k)}s^{k-1}e^{-k \mu s}$ for $s \geq 0$. For this distribution, $\mathbb{E}[S]=\frac{1}{\mu}$ and $k >0$ is the shape parameter that determines the variance and tail behavior. As $k$ gets smaller, the tail gets heavier. We have the following closed form expressions for Gamma distribution:
\begin{align*}
MGF^{(S)}_{\lambda}&=\left(1+\frac{\lambda }{k \mu}\right)^{-k}, \ MGF^{(S,1)}_{\lambda}=\frac{1}{\mu}\left(1+\frac{\lambda }{k \mu}\right)^{-k-1} \\
MGF^{(S,2)}_{\lambda}&=\frac{k+1}{k \mu}\left(1+\frac{\lambda }{k \mu}\right)^{-k-2}
\end{align*}

\begin{figure}[!t]
\centering{
%\hspace{-0.3cm} 
\includegraphics[totalheight=0.35\textheight]{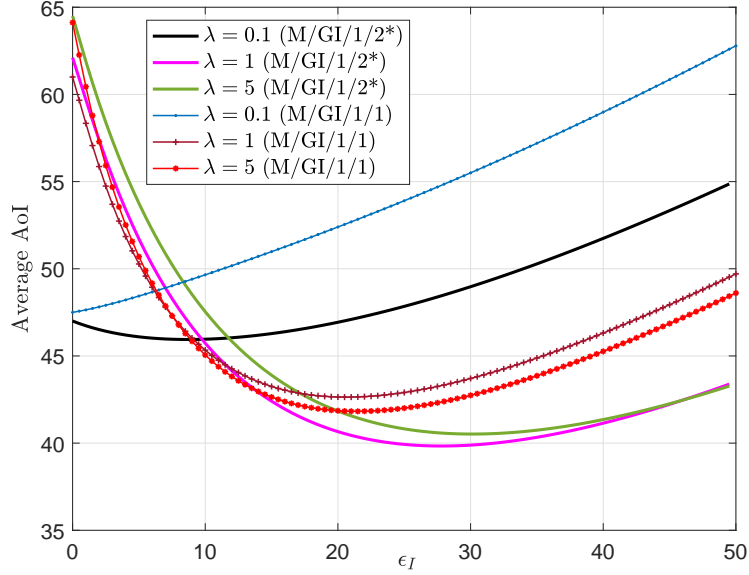}}\vspace{-0.1in}
\caption{\sl For both M/GI/1/1 and M/GI/1/$2^*$ schemes, the plot shows average AoI versus $\epsilon_{I}$ for Gamma distributed service time with $\mu=0.1$, $\epsilon_{B}=0$ (for M/GI/1/$2^*$), $k=0.1$ and various $\lambda$ values. }
\label{fig:local221} 
\end{figure}

In Fig. \ref{fig:local221}, we plot average AoI versus $\epsilon_{I}$ for both M/GI/1/1 and M/GI/1/$2^*$ schemes under Gamma distributed service time with $\mu=0.1$, $\epsilon_{B}=0$, $k=0.1$ and various $\lambda$ values. In this particular case, zero waiting coincides with the optimal selection of waiting in both packet management schemes. We observe that longer waiting periods are more useful for larger arrival rates. This is directly related to the fact that longer waiting periods enable capturing newer arrivals and available service rate is made more efficient use with waiting. Optimal deterministic waiting enables a significant drop in average AoI, especially for larger $\lambda$ values. These observations are similar in the inverse Gaussian distributed service and therefore we show them just for Gamma distributed service. 

\begin{figure}[t] 
%\vspace{-3mm}
    \centering
\subfigure[]
    {
    \includegraphics[totalheight=0.35\textheight]{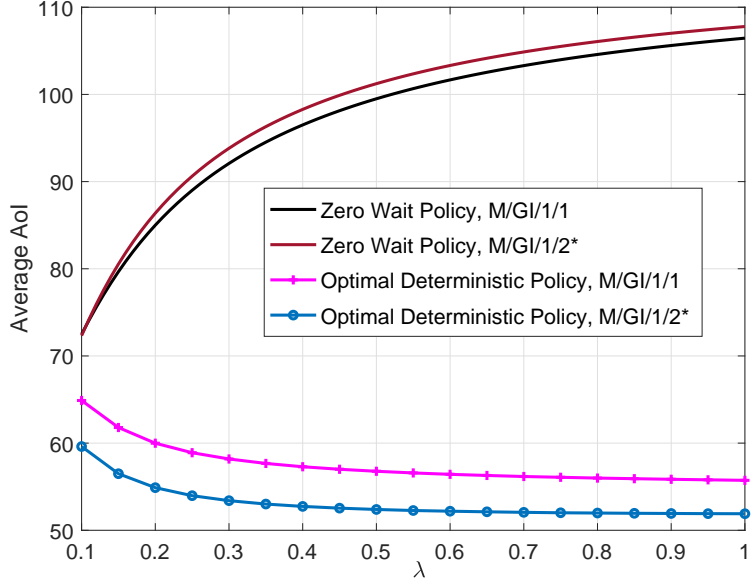}
    \label{fig:local22a}} 
\subfigure[]
    {
    \includegraphics[totalheight=0.35\textheight]{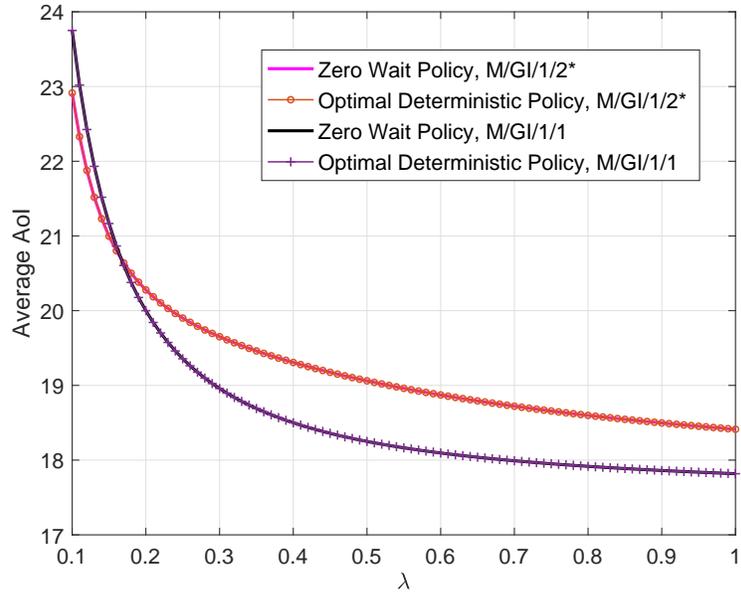}
    \label{fig:local22b}} 
    \vspace{-0mm} 
 \caption{\sl The plots are for average AoI versus arrival rate $\lambda$ under Gamma service distribution with $\mu=0.1$ and show the comparison of zero waiting versus optimal deterministic waiting under M/GI/1/1 and M/GI/1/$2^*$ schemes. (a) The Gamma distribution parameter is $k=0.05$  (b) The Gamma distribution parameter is $k=2$.  \vspace{-3mm}}
  \label{fig:local222} 
\end{figure} 

In Fig. \ref{fig:local222}, we compare the average AoI with zero-waiting and optimal deterministic waiting for both M/GI/1/1 and M/GI/1/$2^*$ schemes with respect to arrival rate $\lambda$ under two different parameters $k$ for Gamma distribution ($k=0.05$ and $k=2$) and fixed service rate $\mu=0.1$. In Fig. \ref{fig:local22a}, it is seen that M/GI/1/1 performs better with respect to M/GI/1/$2^*$ under zero waiting whereas this order is reversed under optimal deterministic waiting. We observe that as the variance of the service distribution is increased (i.e., $k$ is decreased), average AoI increases and the percent improvement in AoI brought by deterministic waiting also increases. In general, as the variance increases, the tail of the service distribution gets heavier and hence this observation supports \cite{sun2017update} in our queueing system. It is seen that this improvement is more dramatic as $\lambda$ increases. We also observe that exponential and Erlang type service distributions (which correspond to $k=1$ and $k=2$) promise little improvement in average AoI by using a deterministic waiting strategy. In particular, Fig. \ref{fig:local22b} reveals that zero wait policy is optimal among all deterministic waiting policies for all $\lambda$ when $k=2$ is assumed. 

\begin{figure}[!t]
\centering{
%\hspace{-0.3cm} 
\includegraphics[totalheight=0.35\textheight]{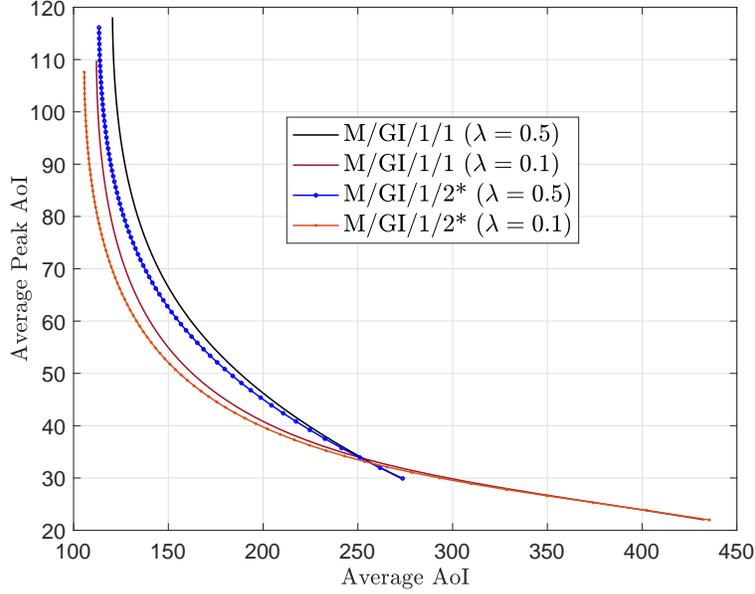}}\vspace{-0.1in}
\caption{\sl The tradeoff curves between average AoI and average peak AoI for M/GI/1/1 and M/GI/1/$2^*$ schemes.}
\label{fig:local5} 
\end{figure}

In Fig. \ref{fig:local5}, we observe the tradeoff curves for achieved average AoI versus achieved average peak AoI under both packet management schemes. The service rate is $\mu=0.1$ and shape parameter is $k=0.01$.We observe that M/GI/1/1 achieves a better tradeoff with respect to M/GI/1/$2^*$ for both $\lambda=0.1$ and $\lambda=0.5$. These curves reveal the cost paid in terms of average peak AoI while minimizing average AoI through introducing waiting before service starts.

\section{Conclusions}
\label{sec:Conc}

In this paper, we investigate Age of Information (AoI) in average and average peak senses for M/GI/1/1 and M/GI/1/$2^*$ queueing disciplines with deterministic waiting deliberately introduced before service starts. Waiting is known to booth average AoI~\cite{sun2017update} especially when preemption is not an option. In this paper, we apply waiting when packets arrive randomly from an outside source and no feedback of the server state is available. Depending on the system state, the server waits a deterministic time before starting service of the packet in the queue. We determine average AoI and average peak AoI expressions for aforementioned queuing disciplines. Our numerical results demonstrate the benefits of waiting in the average AoI under inverse Gaussian and Gamma (exponential and Erlang as special cases) distributed service times. We observe the improvement in average AoI comes at the expense of increased average peak AoI. Our numerical results show that waiting is especially helpful for heavy-tailed service distributions such as inverse Gaussian distribution while the improvement is limited for light-tailed ones such as exponential and Erlang distributions. With regard to the comparison between M/GI/1/1 and M/GI/1/$2^*$, our results show that  the latter outperforms the former once waiting is optimized. A useful byproduct of our analysis is a new method of partitioning the area under AoI evolution curve to calculate time average AoI, yielding a direct connection of the average AoI expression for M/GI/1/1 and M/GI/1/$2^*$ packet management schemes to the expressions for FCFS queuing discipline \cite{kaul2012real}.

\appendix

\subsection{$\mathbb{E}[X_i T_i | \mbox{(I)}]$ for M/GI/1/1}\label{app:01}

\subsubsection{$\epsilon_{I} - X_i \leq 0, \epsilon_{I} - X_i + S_{i-1} > 0$}

In this case, $T_i = \epsilon_{I}-X_i+S_{i-1}+\tilde{X}_i+\epsilon_{I}+S_i$ where $\tilde{X}_i$ is an independent exponentially distributed random variable with mean $\frac{1}{\lambda}$ which represents the arrival time the packet that comes after the service is finished. Let the event $E_1$ denote $\epsilon_{I} - X_i \leq 0, \epsilon_{I} - X_i + S_{i-1} > 0$. We have

\begin{align}\nonumber
&\mathbb{E}[X_iT_i | (I), E_1]\mbox{Pr}[E_1]= \mathbb{E}[X_i\left(\epsilon_{I}-X_i+S_{i-1}+\tilde{X}_i+\epsilon_{I}+S_i\right) | E_1]\mbox{Pr}[E_1]  \\ \nonumber &=(\mathbb{E}[S]+2\epsilon_{I}+\frac{1}{\lambda})e^{-\lambda\epsilon_{I}} \frac{1+\lambda \epsilon_{I}}{\lambda} -\frac{e^{-\lambda\epsilon_{I}}}{\lambda^2}(2+2\lambda\epsilon_{I}+\lambda^2\epsilon_{I}^2) \\ \nonumber &\quad -(\mathbb{E}[S]+2\epsilon_{I}+\frac{1}{\lambda})\frac{e^{-\lambda\epsilon_{I}}}{\lambda}((1+\lambda\epsilon_{I})MGF^{(S)}_{\lambda}+\lambda MGF^{(S,1)}_{\lambda}) \\
&-\frac{e^{-\lambda \epsilon_{I}}}{\lambda}((1+\lambda\epsilon_{I})MGF_{\lambda}^{(S,1)}+\lambda MGF^{(S,2)}_{\lambda}) +\mathbb{E}[S]\frac{e^{-\lambda\epsilon_{I}}}{\lambda}(1+\lambda\epsilon_{I})\nonumber \\ &+\frac{e^{-\lambda\epsilon_{I}}}{\lambda^2}((2+2\lambda \epsilon_{I}+\lambda^2\epsilon_{I}^2)MGF^{(S)}_{\lambda} +(2\lambda+2\lambda^2\epsilon_{I})MGF^{(S,1)}_{\lambda}+\lambda^2MGF^{(S,2)}_{\lambda}) \nonumber
\end{align}

\subsubsection{$\epsilon_{I} > X_i$}

In this case, $T_i = \epsilon_{I}-X_i+S_i$. Let the event $E_2$ denote $\epsilon_{I} > X_i$. We have
\begin{align}\nonumber
&\mathbb{E}[X_iT_i | (I), E_2]\mbox{Pr}[E_2]= \int_0^{\epsilon_{I}} x\left(\epsilon_{I}-x+\mathbb{E}[S]\right)\lambda e^{-\lambda x}dx \\ \nonumber &= \frac{\epsilon_{I}+\mathbb{E}[S]}{\lambda}\left(1-e^{-\lambda \epsilon_{I}}(1+\lambda \epsilon_{I})\right) -\frac{1}{\lambda^2}\left(2-e^{-\lambda \epsilon_{I}}(2+2\lambda \epsilon_{I}+\lambda^2 \epsilon_{I}^2)\right)\nonumber
\end{align}

\subsubsection{$\epsilon_{I} + S_{i-1} \leq X_i$}

In this case, $T_i = \epsilon_{I}+S_{i}$. Let the event $E_3$ denote $\epsilon_{I} + S_{i-1} \leq X_i$. We have
\begin{align}\nonumber
&\mathbb{E}[X_iT_i | (I), E_3]\mbox{Pr}[E_3]= \mathbb{E}[X_i\left(\epsilon_{I}+S_i\right) | E_3]\mbox{Pr}[E_3] \\ &= \left(\epsilon_{I}+\mathbb{E}[S]\right)e^{-\lambda \epsilon_{I}}\left((\epsilon_{I}+\frac{1}{\lambda})MGF_{\lambda}^{(S)}+MGF_{\lambda}^{(S,1)}\right) \nonumber
\end{align}
We finally sum the three expressions to get $\mathbb{E}[X_i T_i | (I)]$:
\begin{align}
\mathbb{E}[X_i T_i | (I)]=\sum_{i=1}^3E[X_i T_i | (I),E_i]\mbox{Pr}[E_i]
\end{align}

\vspace{0.5in}

\subsection{$\mathbb{E}[X_i T_i | \mbox{(W)}]$ for M/GI/1/1}
\label{app:02}

Let $g_{1/1}(r)$ be defined as $\mathbb{E}[X_iT_i | R_{i-1}=r]$ where $R_{i-1}$ denotes residual time for packet $i-1$ to start service. It is expressed as:

\begin{align}\nonumber
g_{1/1}(r)&= \frac{r+\mathbb{E}[S]}{\lambda}\left(1-e^{-\lambda r}(1+\lambda r)\right) \\ \nonumber&\qquad -\frac{1}{\lambda^2}\left(2-e^{-\lambda r}(2+2\lambda r+\lambda^2 r^2)\right) \\ \nonumber &+(\mathbb{E}[S]+r+\epsilon_{I}+\frac{1}{\lambda})e^{-\lambda r} \frac{1+\lambda r}{\lambda} -\frac{e^{-\lambda r}}{\lambda^2}(2+2\lambda r+\lambda^2 r^2) \\ \nonumber &-(\mathbb{E}[S]+\epsilon_{I} + \frac{1}{\lambda}+r)\frac{e^{-\lambda r}}{\lambda} ((1+\lambda r)MGF^{(S)}_{\lambda}+\lambda MGF^{(S,1)}_{\lambda}) \\ &-\frac{e^{-\lambda r}}{\lambda}((1+\lambda r)MGF_{\lambda}^{S,1}+\lambda MGF^{(S,2)}_{\lambda}) \nonumber \\ &+\mathbb{E}[S]\frac{e^{-\lambda r}}{\lambda}(1+\lambda r)+\frac{e^{-\lambda r}}{\lambda^2}((2+2\lambda r+\lambda^2 r^2)MGF^{(S)}_{\lambda} \nonumber \\ &+(2\lambda+2\lambda^2 r)MGF^{(S,1)}_{\lambda}+\lambda^2MGF^{(S,2)}_{\lambda}) \nonumber  \\ &+\left(\epsilon_{I}+ \mathbb{E}[S]\right)e^{-\lambda r}\left((r+\frac{1}{\lambda})MGF_{\lambda}^{(S)}+MGF_{\lambda}^{(S,1)}\right) \nonumber
\end{align}  
where $r$ represents an arbitrary residual time $r \in [0,\epsilon_{I}]$. We have $g_{1/1}(\epsilon_{I})=\mathbb{E}[X_i T_i | (I)]$ and moreover $\mathbb{E}[X_i T_i | (W)]=\frac{1}{\epsilon_{I}}\int_0^{\epsilon_{I}}g(r)dr$. This indefinite integral is

\begin{align}\nonumber
&h_{1/1}(x)=\frac{1}{\lambda^2}\Bigg(e^{-\lambda x}(2\mathbb{E}[S] - \frac{3}{\lambda}) + (\mathbb{E}[S]\lambda - 2)x \\ &\quad + 0.5\lambda x^2 + xe^{-\lambda x}(\mathbb{E}[S]\lambda - 1)\nonumber \\
&\quad-e^{-\lambda x}(2\epsilon_{I} + 2\mathbb{E}[S] - \frac{1}{\lambda}) - xe^{-\lambda x}((\epsilon_{I}+\frac{1}{\lambda})\lambda + \mathbb{E}[S]\lambda - 1) \nonumber \\ \nonumber &\quad+\lambda e^{-\lambda x}(MGF_{\lambda}^{(S,1)}( \epsilon_{I} + \frac{1}{\lambda} + \mathbb{E}[S] + x) \\ \nonumber &\quad+ MGF_{\lambda}^{(S)}(x^2 + (\epsilon_{I} + \frac{1}{\lambda}) x + \mathbb{E}[S] x)) + \frac{3MGF_{\lambda}^{(S)}e^{-\lambda x}}{\lambda} \\ \nonumber &\quad+ e^{-\lambda x}(MGF_{\lambda}^{(S,1)} + 2MGF_{\lambda}^{(S)}(\epsilon_{I} + \frac{1}{\lambda} + \mathbb{E}[S] + 3x)) \\ \nonumber&\quad+e^{-\lambda x}(MGF_{\lambda}^{(S,1)}(2+\lambda x) + MGF_{\lambda}^{(S,2)}\lambda-\mathbb{E}[S](\lambda x +2)) \\ &\quad-\frac{e^{-\lambda x}}{\lambda}(MGF_{\lambda}^{(S)}\lambda^2x^2 + 2MGF_{\lambda}^{(S,1)}\lambda^2x + MGF_{\lambda}^{(S,2)}\lambda^2 \nonumber \\ &\quad+ 4MGF_{\lambda}^{(S)}\lambda x + 4MGF_{\lambda}^{(S,1)}\lambda + 6MGF_{\lambda}^{(S)}) \nonumber \\ \nonumber &\quad-e^{-\lambda x}(\epsilon_{I} + \mathbb{E}[S])(MGF_{\lambda}^{(S)}(2+\lambda x) + MGF_{\lambda}^{(S,1)}\lambda) \Bigg) \nonumber
\end{align}
and $\mathbb{E}[X_i T_i | (W)]=\frac{h_{1/1}(\epsilon_{I}) - h_{1/1}(0)}{\epsilon_{I}}$.

\subsection{$\mathbb{E}[X_i T_i | \mbox{(B)}]$ for M/GI/1/1}\label{app:03}

In case $R_{i-1} \geq X_i$, we have $T_i = R_{i-1}-X_i+\tilde{X}_i + \epsilon_{I}+S_{i}$ where $\tilde{X}_i$ represents the additional waiting time for a new arrival after the service ends. $\tilde{X}_i$ is an independent exponentially distributed random variable with mean $\frac{1}{\lambda}$. Let the event $E_5$ denote $R_{i-1} \geq X_i$. We have
\begin{align}\nonumber
&\hspace{-0.2in}\mathbb{E}[X_iT_i | (B), E_5]\mbox{Pr}[E_5]= \mathbb{E}[X_i\left(R_{i-1}-X_i + \tilde{X}_i +\epsilon_{I}+S_{i}\right) | E_5]\mbox{Pr}[E_5] \\ \nonumber &\hspace{-0.2in}= \frac{\lambda(\epsilon_{I}+\mathbb{E}[S]+\mathbb{E}[R])-1}{\lambda^2}+\frac{MGF^{(R)}_{\lambda}(1-\lambda(\epsilon_{I} +\mathbb{E}[S]))}{\lambda^2} -(\epsilon_{I}+\mathbb{E}[S])MGF^{(R,1)}_{\lambda} 
\end{align}

\subsection{$\mathbb{E}[X_i T_i | \mbox{(I)}]$ for M/GI/1/$2^*$}\label{app:1}

\subsubsection{$\epsilon_{I} - X_i \leq 0, \epsilon_{I} - X_i + S_{i-1} > 0$}

In this case, $T_i = \epsilon_{I}-X_i+S_{i-1}+\epsilon_{B}+S_i$. Let the event $\widehat{E}_1$ denote $\epsilon_{I} - X_i \leq 0, \epsilon_{I} - X_i + S_{i-1} > 0$. We have
\begin{align}\nonumber
&\mathbb{E}[X_iT_i | (I), \widehat{E}_1]\mbox{Pr}[\widehat{E}_1]= \mathbb{E}[X_i\left(\epsilon_{I}-X_i+S_{i-1}+\epsilon_{B}+S_i\right) | \widehat{E}_1]\mbox{Pr}[\widehat{E}_1]  \\ \nonumber &=(2\mathbb{E}[S]+\epsilon_{I}+\epsilon_{B})e^{-\lambda\epsilon_{I}} \frac{1+\lambda \epsilon_{I}}{\lambda} -\frac{e^{-\lambda\epsilon_{I}}}{\lambda^2}(2+2\lambda\epsilon_{I}+\lambda^2\epsilon_{I}^2) \\ \nonumber &\quad -(\mathbb{E}[S]+\epsilon_{B}+\epsilon_{I})\frac{e^{-\lambda\epsilon_{I}}}{\lambda}((1+\lambda\epsilon_{I})MGF^{(S)}_{\lambda}+\lambda MGF^{(S,1)}_{\lambda}) \\
&-\frac{e^{-\lambda \epsilon_{I}}}{\lambda}((1+\lambda\epsilon_{I})MGF_{\lambda}^{(S,1)}+\lambda MGF^{(S,2)}_{\lambda}) \nonumber +\frac{e^{-\lambda\epsilon_{I}}}{\lambda^2}((2+2\lambda \epsilon_{I}+\lambda^2\epsilon_{I}^2)MGF^{(S)}_{\lambda} \\ &+(2\lambda+2\lambda^2\epsilon_{I})MGF^{(S,1)}_{\lambda}+\lambda^2MGF^{(S,2)}_{\lambda}) \nonumber
\end{align}

\subsubsection{$\epsilon_{I} > X_i$}

In this case, $T_i = \epsilon_{I}-X_i+S_i$. Let the event $\widehat{E}_2$ denote $\epsilon_{I} > X_i$. We have
\begin{align}\nonumber
&\mathbb{E}[X_iT_i | (I), \widehat{E}_2]\mbox{Pr}[\widehat{E}_2]= \int_0^{\epsilon_{I}} x\left(\epsilon_{I}-x+\mathbb{E}[S]\right)\lambda e^{-\lambda x}dx \\ \nonumber &= \frac{\epsilon_{I}+\mathbb{E}[S]}{\lambda}\left(1-e^{-\lambda \epsilon_{I}}(1+\lambda \epsilon_{I})\right) -\frac{1}{\lambda^2}\left(2-e^{-\lambda \epsilon_{I}}(2+2\lambda \epsilon_{I}+\lambda^2 \epsilon_{I}^2)\right)\nonumber
\end{align}

\subsubsection{$\epsilon_{I} + S_{i-1} \leq X_i$}

In this case, $T_i = \epsilon_{I}+S_{i}$. Let the event $\widehat{E}_3$ denote $\epsilon_{I} + S_{i-1} \leq X_i$. We have
\begin{align}\nonumber
&\mathbb{E}[X_iT_i | (I), \widehat{E}_3]\mbox{Pr}[\widehat{E}_3]= \mathbb{E}[X_i\left(\epsilon_{I}+S_i\right) | \widehat{E}_3]\mbox{Pr}[\widehat{E}_3] \\ &= \left(\epsilon_{I}+\mathbb{E}[S]\right)e^{-\lambda \epsilon_{I}}\left((\epsilon_{I}+\frac{1}{\lambda})MGF_{\lambda}^{(S)}+MGF_{\lambda}^{(S,1)}\right) \nonumber
\end{align}
We finally sum the three expressions to get $\mathbb{E}[X_i T_i | (I)]$:
\begin{align}
\mathbb{E}[X_i T_i | (I)]=\sum_{i=1}^3E[X_i T_i | (I),\widehat{E}_i]\mbox{Pr}[\widehat{E}_i]
\end{align}

\subsection{Definition of $g(r)$ for M/GI/1/$2^*$}\label{app:2}

$g_{1/2^*}(r)$ defines $\mathbb{E}[X_iT_i | R_{i-1}=r]$ where $R_{i-1}$ denotes residual time for packet $i-1$ to start service. It is expressed as:

\begin{align}\nonumber
g_{1/2^*}(r)&= \frac{r+\mathbb{E}[S]}{\lambda}\left(1-e^{-\lambda r}(1+\lambda r)\right) -\frac{1}{\lambda^2}\left(2-e^{-\lambda r}(2+2\lambda r+\lambda^2 r^2)\right) \\ \nonumber &+(2\mathbb{E}[S]+r+\epsilon_{B})e^{-\lambda r} \frac{1+\lambda r}{\lambda} -\frac{e^{-\lambda r}}{\lambda^2}(2+2\lambda r+\lambda^2 r^2) \\ \nonumber &-(\mathbb{E}[S]+\epsilon_{B}+r)\frac{e^{-\lambda r}}{\lambda} ((1+\lambda r)MGF^{(S)}_{\lambda}+\lambda MGF^{(S,1)}_{\lambda}) \\ &-\frac{e^{-\lambda r}}{\lambda}((1+\lambda r)MGF_{\lambda}^{S,1}+\lambda MGF^{(S,2)}_{\lambda}) \nonumber \\ &+\frac{e^{-\lambda r}}{\lambda^2}((2+2\lambda r+\lambda^2 r^2)MGF^{(S)}_{\lambda} \nonumber \\ &+(2\lambda+2\lambda^2 r)MGF^{(S,1)}_{\lambda}+\lambda^2MGF^{(S,2)}_{\lambda}) \nonumber  \\ &+\left(\epsilon_{I}+\mathbb{E}[S]\right)e^{-\lambda r}\left((r+\frac{1}{\lambda})MGF_{\lambda}^{(S)}+MGF_{\lambda}^{(S,1)}\right) \nonumber
\end{align}  
where $r$ represents an arbitrary residual time $r \in [0,\epsilon_{I}]$. We have $g_{1/2^*}(\epsilon_{I})=\mathbb{E}[X_i T_i | (I)]$ and moreover
\begin{align}\nonumber
\mathbb{E}[X_i T_i | (WaI)]=\frac{1}{\epsilon_{I}}\int_0^{\epsilon_{I}}g_{1/2^*}(r)dr
\end{align} 

\subsection{$\mathbb{E}[X_i T_i | \mbox{(WaI)}]$ and $\mathbb{E}[X_i T_i | \mbox{(WaB)}]$ for M/GI/1/$2^*$}\label{app:4}

The indefinite integral in (\ref{eq:exp1}) is\vspace{-0.05in}
\begin{align}\nonumber
&h_{1/2^*}(x)=\frac{1}{\lambda^2}\Bigg(e^{-\lambda x}(2\mathbb{E}[S] - \frac{3}{\lambda}) + (\mathbb{E}[S]\lambda - 2)x \\ &\quad + 0.5\lambda x^2 + xe^{-\lambda x}(\mathbb{E}[S]\lambda - 1)\nonumber \\
&\quad-e^{-\lambda x}(2\epsilon_{B} + 2\mathbb{E}[S] - \frac{3}{\lambda}) - xe^{-\lambda x}(\epsilon_{B}\lambda + \mathbb{E}[S]\lambda - 1) \nonumber \\ \nonumber &\quad+\lambda e^{-\lambda x}(MGF_{\lambda}^{(S,1)}( \epsilon_{B} + \mathbb{E}[S] + x) \\ \nonumber &\quad+ MGF_{\lambda}^{(S)}(x^2 + \epsilon_{B} x + \mathbb{E}[S] x)) + \frac{3MGF_{\lambda}^{(S)}e^{-\lambda x}}{\lambda} \\ \nonumber &\quad+ e^{-\lambda x}(MGF_{\lambda}^{(S,1)} + 2MGF_{\lambda}^{(S)}(\epsilon_{B} + \mathbb{E}[S] + 3x)) \\ \nonumber&\quad+e^{-\lambda x}(MGF_{\lambda}^{(S,1)}(2+\lambda x) + MGF_{\lambda}^{(S,2)}\lambda-\mathbb{E}[S](\lambda x +2)) \\ &\quad- \frac{e^{-\lambda x}}{\lambda}(MGF_{\lambda}^{(S)}\lambda^2x^2 + 2MGF_{\lambda}^{(S,1)}\lambda^2x + MGF_{\lambda}^{(S,2)}\lambda^2 \nonumber \\ &\quad+ 4MGF_{\lambda}^{(S)}\lambda x + 4MGF_{\lambda}^{(S,1)}\lambda + 6MGF_{\lambda}^{(S)}) \nonumber \\ \nonumber &\quad-e^{-\lambda x}(\epsilon_{I} + \mathbb{E}[S])(MGF_{\lambda}^{(S)}(2+\lambda x) + MGF_{\lambda}^{(S,1)}\lambda) \Bigg) \nonumber
\end{align}
Then, we have $\mathbb{E}[X_i T_i | (WaI)]=\frac{h_{1/2^*}(\epsilon_I)-h_{1/2^*}(0)}{\epsilon_I}$ and
$\mathbb{E}[X_i T_i | (WaB)]=\frac{h_{1/2^*}(\epsilon_B)-h_{1/2^*}(0)}{\epsilon_B}$.

\subsection{$\mathbb{E}[X_i T_i | \mbox{(B)}]$ for M/GI/1/$2^*$}\label{app:3}

\subsubsection{$R_{i-1} < X_i$}
In this case, the event $\widehat{E}_4$ denotes $R_{i-1} < X_i$ and replacing $\epsilon_I$ with $\epsilon_B$ in cases 1 and 2 in Appendix \ref{app:1} as well as decomposing $X_i$ as $X_i-R_{i-1}$ plus $R_{i-1}$ in all cases considered in Appendix \ref{app:1}:
\begin{align}\nonumber
\hspace{-0.2in}\mathbb{E}[R_{i-1} T_i | (B), \widehat{E}_4]\mbox{Pr}[\widehat{E}_4] &= MGF_{\lambda}^{(R,1)}((\epsilon_{B}+\mathbb{E}[S]-\frac{1}{\lambda})+\mathbb{E}[S] +(\frac{1}{\lambda}+\epsilon_{I}-\epsilon_{B})MGF^{(S)}_{\lambda})e^{-\lambda\epsilon_{B}} \nonumber \\ \nonumber &\quad +MGF_{\lambda}^{(R,1)}((\epsilon_{B}+\mathbb{E}[S])(1-e^{-\lambda\epsilon_{B}})-\frac{1}{\lambda} +\frac{e^{-\lambda\epsilon_{B}}}{\lambda}(1+\lambda\epsilon_{B}))
\end{align}

\subsubsection{$R_{i-1} \geq X_i$}

In this case, $T_i = R_{i-1}-X_i+\epsilon_{B}+S_{i}$. Let the event $\widehat{E}_5$ denote $R_{i-1} \geq X_i$. We have
\begin{align}\nonumber
&\hspace{-0.2in}\mathbb{E}[X_iT_i | (B), \widehat{E}_5]\mbox{Pr}[\widehat{E}_5]= \mathbb{E}[X_i\left(R_{i-1}-X_i+\epsilon_{B}+S_{i}\right) | \widehat{E}_5]\mbox{Pr}[\widehat{E}_5] \\ \nonumber &\hspace{-0.2in}= \frac{\lambda(\epsilon_{B}+\mathbb{E}[S]+\mathbb{E}[R])-2}{\lambda^2}+\frac{MGF^{(R)}_{\lambda}(2-\lambda(\epsilon_{B}+\mathbb{E}[S]))}{\lambda^2} +(1-\lambda \epsilon_{B}-\lambda \mathbb{E}[S])\frac{MGF^{(R,1)}_{\lambda}}{\lambda} 
\end{align}

\end{document}